\documentstyle[preprint,multicol,aps,epsf]{revtex} 
\begin{document}
\title{Topological Spin Pumps}
\author{Fei Zhou}
\address{Department of Physics and Astronomy, University of British 
Columbia,\\
6224 Agriculture Road, Vancouver, B.C.V6T 1Z1,Canada}
\date{\today}
\maketitle

\begin{abstract} 
We have established semiclassical kinetic equations for various spin correlated
pumping phenomena incorporating {\em adiabatic} spin rotation in wave 
functions. We employ this technique to study
topological pumps and illustrate spin pumping in a few models where various spin configurations or
{\em topological motors} drive adiabatic pumps.  In the Rashba model we find that a topological spin
pump is driven by a meron with {\em positive} one half skyrmion charge, the size of which can be
controlled by external applied gates or Zeeman fields.  In the Dresselhaus model on the other hand,
electron spins are pumped out by a {\em negative} meron. We examine the effects of Zeeman fields on
topological spin pumping and responses of Fermi seas in various topological pumps. The phenomena of
topological pumping are attributed to the beam splitting of electrons in the presence of spin rotation, or
{\em topological Stern-Gerlach splitting} and occur in a transverse direction along which charge
pumping currents might either vanish or are negligible. 
The transport equations established here might also
be applied to the studies of anomalous Hall effect and spin Hall effect as demonstrated in one of
the appendices.  
All results are obtained in an adiabatic expansion where the {\em adiabaticity} conditions in Eqs.(\ref{ad1}),(\ref{ad2}),(\ref{ad4}),(\ref{ad5}) are 
satisfied.
\end{abstract}

\section{Introduction}

Adiabatic transport of electrons or quantum pumping 
which is nearly reversible has been one of very promising means 
to manipulate coherent wave packets in the extreme quantum limit.
In the presence of periodic adiabatic perturbations,
a net charge can be transferred across a quantum structure during each 
period which is independent of frequencies of external perturbations and 
which represents a DC current 
induced by adiabatic perturbations.
This phenomenon was first observed in an early work on 
transport of edge electrons in quantum Hall states
\cite{Laughlin81}. But general solutions to this problem 
were provided in Ref.\cite{Thouless83} where conditions 
of quantized charge transport were established.
The robustness of quantized transport with respect to disorder
potentials and applications to quantum Hall effects 
were later on studied in a series of works\cite{Niu84}.

The absence of dissipation during the adiabatic process is evident if
perturbations are applied to a closed quantum structure with nontrivial
topology such as a mesoscopic ring or torus. If one further assumes that the 
electron
spectrum is discrete and the external frequency is incommensurate 
with energy gaps in the spectrum, then no resonance absorption can occur
and quantum states evolve via unitary transformation which 
conserves the entropy.
The absence of entropy production in the adiabatic process is therefore a 
natural consequency of a pure state evolution which has been known for 
a while.

For a quantum structure with continuous spectra either because
of contact with leads in a mesoscopic limit or more generally because of 
thermal
broadening, it is more convenient to introduce one-particle density
matrices to describe the evolution of quantum systems. The adiabaticity can be achieved when
the frequency of applied perturbations is lower than various relaxation
rates characterizing the dynamics of one-particle density matrix.
The issue of entropy
production in this case however hasn't been fully addressed and is less well understood. 
Nevertheless, it is widely appreciated that
dissipation involved in adiabatic transport in this limit should 
also be smaller than 
that due to a transport current with biased voltages applied.

Given the obvious advantage of the adiabatic charge transport, in this 
article I 
intend to generalize the idea to adiabatic spin transport.
I will address the issue of spin pumping in both
limits which can be easily achieved in laboratories
and limits which are theoretically exciting but might not be
as easy to be realized in solid state structures.
Particularly, I will propose a novel spin pumping mechanism which is
based on the topological beam splitting of electrons
instead of the usual Zeeman splitting.
Special classes of models are introduced to facilitate discussions
on topological spin pumping.
At the end, I will compare the efficiency of different spin pumping
schemes.

One obvious mean to pump spin out of the system is 
to adiabatically transfer polarized electrons in quantum structures.
During the adiabatic transport, the currents carried by spin-up and spin-down 
electrons have an asymmetric part and therefore 
electrons pumped out of the structures also carry net spins.
This standard scheme is reviewed in section III.

In section IV, I discuss the phenomena of topological beam splitting
in details. Particularly, I demonstrate that spin rotation in either
real-space ($X$-space) or in Fermi seas leads to transverse motion of electrons. 
In section V and VI, I investigate topological spin pumping
due to spin rotation either in the $X$-space or in Fermi seas. In both cases, 
spin-up and spin-down electrons
though both are electrically negatively charged, carry certain topological 
charges with
opposite signs. As consequencies,  
spin-up and spin-down 
electrons can be split because of opposite topological transverse 
forces,
an analogy of splitting of electrons and positrons in an orbital magnetic
field. I would like to refer this kind of beam splitting as topological 
Stern-Gerlach splitting ({\em TSGS}) to contrast the usual Stern-Gerlach splitting of 
spin-up and -down particles in atomic physics.
In this article I focus on the origin of this phenomenon and
basics features. I plan to present a practical design of a topological 
spin pump in a subsequent paper.

Finally, in connection with topological spin pumping to be
discussed in the article, it is worth mentioning a few recent
works on anomalous Hall effects and spin injection where the topology of
Hilbert spaces plays
a paramount role.
In Ref.\cite{Ye99}, the authors pointed out that interactions between 
conduction
electrons and background skyrmion configurations
activated in magnetite might be responsible for the sign and
temperature dependence of anomalous Hall 
effects observed in Ref.\cite{Matl98}.
The authors of Ref.\cite{Sundaram99} meanwhile argued that ${\bf k}$-space Chern-number 
densities 
also modify the equation of motion for electrons. The corresponding
contributions to the anomalous Hall effect in
ferromagnetic semiconductors were further studied in Ref. 
\cite{Jungwirth02}.
Following these works, it is now believed that
the anomalous Hall effect can be an intrinsic phenomenon.
It is indeed likely to occur when
skew-scattering from impurity atoms is absent
as proposed
by Karplus and Luttinger a while ago\cite{Karplus54}.
In Ref.\cite{Sinova03} the authors have considered 
intrinsic spin Hall currents in semiconductors; many interesting features have been found. 
Related discussions can be also found in Ref.\cite{Onoda02}.

Independently, 
in a series of illuminating works\cite{Murakami03,Murakami03a,Bernevig03}
the authors studied spin injection 
in semiconductors characterized by the
Luttinger Hamiltonian. They have found that singular topological structures in the ${\bf k}$-space as 
well have fascinating effects on accelerated electrons and lead to 
important consequencies on spin injection. In Ref.\cite{Murakami03a}, the 
authors further
pointed out possible connection between transverse spin Hall currents 
and supercurrent in superconductors. 
The issue of dissipation however is still under debate and remains to be fully understood.

\section{Kinetic equations for one-particle density matrix}

Consider the one-particle density matrix
$\rho_{\alpha\beta}({\bf x}',{\bf x};t',t)$.
Subsripts $\alpha,\beta=\pm$ are introduced as spin indices;
later in this article I also introduce $\eta,\xi=1,2,..N$ as
indices in an $N$-dimensional parameter space; $\mu,\nu=x,y,z$ as indices
in the real or momentum spaces.
The evolution of one-particle density matrix is determined by
the following equation

\begin{eqnarray}
[i\frac{\partial }{\partial 
t} +i\frac{\partial 
}{\partial 
t'}]
{\rho}_{\alpha\beta}({\bf x}',{\bf x};t',t)
={\cal H}_{\alpha\beta'}({\bf x},i\frac{\partial}{\partial {\bf x}};t) 
\rho_{\beta'\beta}({\bf x}',{\bf x};t',t)
-\rho_{\alpha\beta'}({\bf x}',{\bf x};t',t){\cal H}_{\beta'\beta}({\bf 
x}',i\frac{\partial}{\partial {\bf x}'};t').
\end{eqnarray}

To study the transport in a semiclassical limit, one
introduces
\begin{eqnarray}
&& {\bf r}=\frac{{\bf x}+{\bf x}'}{2}, {\bf X}={\bf x}'-{\bf x};
\nonumber \\
&&T=\frac{t+t'}{2}, \tau=t'-t.
\end{eqnarray}
Furthermore,
one defines a generalized semiclassical density matrix
\begin{eqnarray}
&& \rho_{\alpha\beta}({\bf k},{\bf r};\omega, T)
=\frac{1}{V}\int d{\bf X} d\tau 
\exp(i{\bf k}\cdot
{\bf X}-i\omega \tau) 
\rho_{\alpha\beta}
({\bf r}+\frac{{\bf X}}{2},{\bf r}-\frac{{\bf X}}{2};
T+\frac{\tau}{2},T-\frac{\tau}{2})
\end{eqnarray}
$V$ is the volume of systems.

In a semiclassical approximation, one obtains the equation of motion for
the one-particle density matrix,

\begin{eqnarray}
&& \frac{\partial \rho_{\alpha\beta}({\bf k},{\bf r};\omega,T)}{\partial 
T}
+[\frac{\partial H_{\alpha\beta'}({\bf k},{\bf r};T)}{\partial {\bf k}}
\frac{\partial}{\partial {\bf r}} 
-\frac{\partial H_{\alpha\beta'}({\bf k},{\bf r};T)}{\partial {\bf r}}
 \frac{\partial}{\partial {\bf k}} 
 +\frac{\partial H_{\alpha\beta'}({\bf k},{\bf r};T)}{\partial T}
\frac{\partial}{\partial \omega}]\otimes 
\rho_{\beta'\beta}({\bf k},{\bf r};\omega,T)\nonumber\\
&&
=\frac{1}{i}H_{\alpha\beta'}({\bf k},{\bf r};T)\tilde{\otimes} 
\rho_{\beta'\beta}
({\bf k},{\bf r};\omega,T)
+{\cal I}^{C.I.}
\rho_{\alpha\beta}({\bf k},{\bf r};\omega,T).
\label{ke1}
\end{eqnarray}
Here 

\begin{eqnarray}
&& A_{\alpha\beta} \otimes B_{\beta\gamma}=
\frac{1}{2}[A_{\alpha\beta}B_{\beta\gamma}
+B_{\alpha\beta}A_{\beta\gamma}],   
\nonumber\\
&& A_{\alpha\beta} \tilde{\otimes} B_{\beta\gamma}=
A_{\alpha\beta}B_{\beta\gamma}
-B_{\alpha\beta}A_{\beta\gamma}.
\end{eqnarray}
and ${\cal I}^{C.I.}$ is a collision integral operator for elastic
(nonmagnetic) scattering processes.
${\bf k}$ and ${\bf r}$ in Eq.\ref{ke1} are variables instead of 
operators.
The gradient expansion which is valid as far as
the transport occurs at a scale much larger than the fermi wave length
is sufficient for the study of semiclassical phenomena.
In all models employed in this article I find
the commutator
$H\tilde{\otimes}\rho$ in Eq.\ref{ke1} vanishes in the semiclassical approximation.

The charge current ${\bf J}$ and spin current ${\bf J}^{z}$ with 
spin along ${\bf 
e}_z$ direction are

\begin{eqnarray}
&& {\bf J}({\bf r}, T)=\int \frac{d\omega}{2\pi} \frac{d^3{\bf 
k}}{(2\pi)^3}
\frac{\partial H_{\alpha\beta}({\bf k},{\bf r})}{\partial {\bf k}}
\rho_{\beta\alpha}({\bf k},{\bf r};\omega,T), \nonumber \\
&& {\bf J}^{z} ({\bf r}, T)=\int \frac{d\omega}{2\pi} \frac{d^3{\bf 
k}}{(2\pi)^3}
{\bf e}_z\cdot {\bf \sigma}_{\alpha\alpha'}
\frac{\partial H_{\alpha'\beta}({\bf k},{\bf r})}{\partial {\bf k}}
\rho_{\beta\alpha}({\bf k},{\bf r};\omega,T). 
\end{eqnarray}
To facilitate discussions on the adiabatic transport,
one further separates the one-particle density matrix into symmetric and
asymmetric parts ($\rho_{\alpha\beta}^{S,A}$),

\begin{eqnarray}
&& \rho_{\alpha\beta}({\bf k},{\bf 
r};\omega,T)=\rho^S_{\alpha\beta}({\bf k},{\bf 
r};\omega,T)+\rho^A_{\alpha\beta}({\bf k},{\bf r};\omega,T)
\nonumber \\
&& {\cal I}^{C.I.}\rho^S_{\alpha\beta}({\bf k},{\bf r};T)=0,
{\cal I}^{C.I.}\rho^A_{\alpha\beta}({\bf k},{\bf r};T)=\frac{1}{\tau_0}
\rho^A_{\alpha\beta}
({\bf k},{\bf r};T).
\end{eqnarray}
And  

\begin{equation}
\int {d{\bf \Omega}({\bf k})} \rho^A_{\alpha\beta}({\bf k},{\bf 
r};\omega,T)
=\int {d{\bf \Omega}}({\bf k}) {\bf \Omega}({\bf k}) 
\rho^S_{\alpha\beta}({\bf k},{\bf 
r};\omega,T)=0,
\label{sa}
\end{equation}
${\bf \Omega}({\bf k})$ is introduced as a unit vector along the direction
of ${\bf k}$ in Eq.\ref{sa} and in the following sections.

In the relaxation approximation employed here, 
elastic {\em nonmagnetic} impurity scattering only leads to momentum
relaxation because the collision integrals are $SU(2)$ singlet operators and 
act trivially on the density matrix $\rho_{\alpha\beta}$.
This however doesn't {\em generally} imply that impurity scattering combined 
with the spin-orbit coupling which I am going to discuss should not cause transitions between 
different spin states. Nevertheless in the adiabatic approximation employed in this article, 
in a special basis these transitions are negligible (see discussions about $SU(2)$ gauge fields and
adiabaticity conditions in section V and VI). 
As far as the adiabaticity conditions are satisfied, the collision integral can be treated in the usual Born approximation
even in the presence of spin-orbit coupling. 
Please see more specific discussions about the adiabaticity at the beginning of section VI, and discussions
after Eqs.\ref{fgf},\ref{ad5}.

Since I am interested in the transport phenomena at distance much longer
than the mean free path $l_0$ or at frequencies much lower than the inverse 
of mean free time $\tau_0$, i.e.,

\begin{equation}
L \gg l_0=\tau_0 v_F, T_0 \gg \tau_0,
\end{equation}
I adopt the standard diffusion approximation.
Furthermore for the study of adiabatic charge and spin pumping phenomena
it is sufficient to keep the first order term in
an adiabatic expansion.
Taking into account the definition of symmetric and antisymmetric components, one 
obtains $\rho^A$ and $\rho^S$ as

\begin{eqnarray}
&&\rho^A_{\alpha\beta}=\tau_0 v_k{\bf \Omega}_{\mu}({\bf k})
[-\frac{\partial}{\partial {\bf r}_{\mu}}\delta_{\alpha\beta'}+\frac{\partial 
H_{\alpha\beta'}({\bf k},{\bf r};T)}{\partial 
{\bf r}_\mu} \otimes
\frac{\partial}{\partial \epsilon_k}]\rho^S_{\beta'\beta}
({\bf k},{\bf r};\omega,T) \\
&& D_k\nabla^2 \rho^S_{\alpha\beta}({\bf k},{\bf 
r};\omega,T)=\frac{\partial H_{\alpha\beta'}({\bf 
k},{\bf r};T)}{\partial T} \otimes \frac{\partial 
\rho^0_{\beta'\beta}({\bf 
k},{\bf r};\omega,T)}
{\partial \omega}.
\label{asy}
\end{eqnarray}
$v_k=|{\bf k}|/m$ is the velocity and $D_k= v_k^2 \tau_0/d$ is a diffusion constant; $d=2,3$ are the dimensions of the Fermi seas which interest us.
$\rho^0_{\alpha\beta}$ is the equilibrium one-particle density matrix.

The charge pumping current and spin pumping current with spin pointing at
the $z$-direction 
can then be expressed as

\begin{eqnarray}
&&{\bf J}_\mu=\int \frac{d\omega}{2\pi}  \frac{d^3{\bf k}}{(2\pi)^3}
D_k [-\frac{\partial}{\partial {\bf 
r}_\mu}\delta_{\alpha\beta}+\frac{\partial H_{\alpha\beta}({\bf 
k}, {\bf r};T)}{\partial 
{\bf r}_\mu} \otimes \frac{\partial} 
{\partial \epsilon_k}]
\rho^S_{\beta\alpha}({\bf k},{\bf r};\omega,T),
\nonumber \\
&& {\bf J}^z_\mu=\int \frac{d\omega}{2\pi} \frac{d^3{\bf 
k}}{(2\pi)^3} D_k {\bf e}_z \cdot {\bf \sigma}_{\alpha\alpha'}
[-\frac{\partial}{\partial {\bf 
r}_\mu}\delta_{\alpha'\beta}+\frac{\partial H_{\alpha'\beta}({\bf 
k}, {\bf r};T)}{\partial 
{\bf r}_\mu}
\otimes
\frac{\partial} 
{\partial \epsilon_k}]
\rho^S_{\beta\alpha}({\bf k},{\bf r};\omega,T).
\nonumber \\
\end{eqnarray}
Here and in the rest of the article I set $\hbar=e=1$.
This set of equation will be used to study various spin pumping phenomena.

\section{Charge and spin pumping of polarized electrons}

I first apply the kinetic equations to study adiabatic charge
transport of polarized electrons. The
external perturbations are represented by $N$ external {\em a.c.} gates
with same period $T_0$; i.e.
\begin{eqnarray}
&& V_{ext}({\bf r},T)=\sum_\eta g_\eta(T) V_\eta(\bf r-{\bf r}_\eta),
\nonumber \\
&& g_\eta(T)=g_\eta(T+T_0),\eta=1,2,...N.
\end{eqnarray}
For pumping phenomena,
the boundary conditions at ${\bf r}=\pm L/2 {\bf e}_\mu$ ($\mu=x,y$) are 
chosen as

\begin{eqnarray}
\rho^S_{\alpha\beta}({\bf k},-\frac{L}{2}{\bf e}_\mu;\omega,T)
=\rho^S_{\alpha\beta}({\bf k},\frac{L}{2}{\bf e}_\mu;\omega,T).
\label{bc}
\end{eqnarray}
Eq.\ref{bc} is valid when a) the sample has a closed geometry along 
${\bf e}_\mu$
or b)more practically leads at boundaries are ideal and are maintained in a thermal 
equilibrium, which also corresponds to a current biased situation.

As noticed in a previous work\cite{Zhou99}, at time $T$ the one-particle density matrix in the presence of
adiabatic perturbation only depends on the potentials at that moment.
Particularly, the density matrix is
a function of ${g_\eta}(T)$, $\eta=1,2,..N$ and their time derivatives $\dot{g}_\eta(T)$; and it 
has this local time dependence as a result of adiabaticity. 
The charge transport per period $T_0$ therefore has the following 
appealing general structure ($\eta,\eta',\xi=1,2,...N$)\cite{Zhou99},

\begin{eqnarray}
Q_{ii}=e\int dg_\eta\wedge dg_\xi \pi_{\eta\xi},
\pi_{\eta\xi}=\big( \frac{\partial}{\partial g_\eta}
\frac{\partial}{\partial \dot{g}_\xi}-\frac{\partial}{\partial g_\xi}
\frac{\partial}{\partial\dot{g}_\eta}\big) Tr\{ \frac{\bf k}{2m L_i} \rho({\bf k},{\bf r};\{g_{\eta'}\}, 
\{\dot{g}_{\eta'}\})\}. 
\label{ctg}
\end{eqnarray}
Here I introduce $Q_{ii}$ as the charge transport along the ${\bf e}_i$-direction.
$dg_\eta\wedge dg_\xi$ is a skew symmetric wedge product, i.e., 
$dg_\eta \wedge dg_\xi$ $=-dg_\xi \wedge dg_\eta$. The trace $Tr$ is carried over the momentum, real-space and spin space.
In Eq.\ref{ctg}, the charge transport has been expressed explicitly in terms of the {\em adiabatic curvature}
$\pi_{\eta\xi}$; the form of the curvature is uniquely defined by the local time-dependence of the one-particle
density matrix $\rho_{\alpha\beta}({\bf k},{\bf r}; \{g_{\eta'}(T)\},\{\dot{g}_{\eta'}(T) \})$.

In the following I am going to evaluate the one-particle density matrix $\rho_{\alpha\beta}
({\bf k},{\bf r};T)$ and therefore the curvature $\pi_{\eta\xi}$ explicitly 
including the spin polarization.
As indicated in Eq.\ref{asy}, the antisymmetric part of the density matrix can be expressed in terms of the symmetric
part. And
the symmetric part of the density matrix receives a nonadiabatic correction following
the second line in Eq.\ref{asy}; the solution is

\begin{eqnarray}
&& \rho^{S}_{\alpha\beta}({\bf k},{\bf r};\omega,T)
=\rho^0_{\alpha\beta}({\bf k};\omega)+
\frac{1}{D_k}M^{S1}_{\alpha\beta}({\bf k},{\bf r};\omega,T),\nonumber \\
&& M^{S1}_{\alpha\beta}({\bf k},{\bf r};\omega,T) =\int d{\bf r}' G({\bf r},{\bf 
r'}) \dot{g}_\eta \frac{\partial}{\partial g_\eta} 
H_{\alpha\beta'}({\bf k},{\bf 
r}'; \{ g_{\eta'}(T) \})
\otimes \frac{\partial \rho^0_{\beta'\beta}({\bf 
k};\omega)}{\partial \omega}.
\label{asy1}
\end{eqnarray}
I have defined
$G({\bf r},{\bf r'})$ as a free propagator

\begin{eqnarray}
&& \nabla^2 G({\bf r},{\bf r'})=\delta({\bf r},{\bf r'}).
\end{eqnarray}
At boundaries, one sets $G({\bf r},{\bf r'})$ to be zero. 
Superscript $S1$ in Eq.\ref{asy1} refers to the first order nonadiabatic corrections
to the symmetry part of one-particle density matrix.
The matrix $M^{S1}_{\alpha\beta}$, 
or more specifically $M^{S1}_{\alpha\beta}({\bf k},{\bf r};\omega,\{ g_{\eta'}(T)\},\{ \dot{g}_{\eta'}(T)\})$
is widely cited in the
rest of this article.

Correspondingly,
one can also calculate the contribution to the asymmetric 
component of density matrix 
in the first order adiabatic approximation
using Eq.10. 
%\begin{eqnarray}
%&&\rho^{A}_{\alpha\beta}=\tau_0 v_F{\bf \Omega}_{k\mu}
%\frac{\partial H_{\alpha\beta'}({\bf k},{\bf X};T)}{\partial 
%X_\mu}\otimes [\frac{\partial \rho^0_{\beta\beta'}({\bf 
%k};\omega)}{\partial \epsilon_k} + 
%G({\bf X},{\bf X'})\frac{\partial H_{\alpha\beta'}({\bf          
%k},{\bf                                                                        
%X}'; T)}                                                                 
%{\partial T} \otimes \frac{\partial^2 \rho^0_{\beta'\beta}({\bf           
%k};\omega)}{\partial \omega\partial \epsilon_k}].
%\end{eqnarray}
Substituting these results into the expression for currents, 
one arrives at
\begin{eqnarray}
&&{\bf J}_\mu=\int \frac{d\omega}{2\pi} \frac{d^3{\bf k}}{(2\pi)^3}
D_k \frac{\partial H_{\alpha\beta}({\bf k},{\bf r};\{ g_{\eta'}(T)\})}{\partial 
{\bf r}_\mu}\otimes 
\frac{\partial }{\partial \epsilon_k}[\frac{1}{D_k}
M^{S1}_{\beta\alpha}({\bf           
k},{\bf r};\omega,\{ g_{\eta'}(T)\}, \{ \dot{g}_{\eta'}(T)\})],
\nonumber\\
%[G({\bf X},{\bf X'})\frac{\partial H_{\beta\gamma}({\bf          
%k},{\bf                                                                        
%X}';\omega,T)}                                                                 
%{\partial T}\otimes 
%\frac{\partial^2 \rho^0_{\gamma\alpha}({\bf           
%k};\omega)}{\partial \omega\partial \epsilon_k}]\nonumber\\
&&{\bf J}^z_\mu=\int \frac{d\omega}{4\pi} \frac{d^3{\bf k}}{(2\pi)^3}
D_k {\bf e}_z \cdot \sigma_{\alpha\alpha'} 
\frac{\partial 
H_{\alpha'\beta}({\bf k},{\bf r};\{ g_{\eta'}(T)\})}{\partial 
{\bf r}_\mu}\otimes 
\frac{\partial}{\partial \epsilon_k}
[\frac{1}{D_k}
M^{S1}_{\beta\alpha}({\bf           
k},{\bf r};\omega,\{ g_{\eta'}(T)\}, \{ \dot{g}_{\eta'}(T) \})].
%[G({\bf X},{\bf X'})\frac{\partial H_{\beta\gamma}
%({\bf k},{\bf  
%X}';\omega,T)}                                                                 
%{\partial T}\otimes \frac{\partial^2 \rho^0_{\gamma\alpha}({\bf           
%k};\omega)}{\partial \omega\partial \epsilon_k}].
\label{cscurrent}
\end{eqnarray}
I have neglected a term which does not contribute to the 
total current because of the boundary conditions in Eq.\ref{bc}.
In the rest of the article, I will use the notion $H_{\alpha\beta}({\bf k}, {\bf r};T)$ and
$M^{S1}_{\alpha\beta}({\bf k},{\bf r};\omega,T)$ without showing $\{ g_{\eta'}(T)\}, \{ \dot{g}_{\eta'}(T)\}$
explicitly.

In the absence of spin-dependent impurity scattering, $S_z$ is a good quantum number;

\begin{eqnarray}
&& \rho^0_{\alpha\beta}({\bf k};\omega)=n_0(\epsilon_{\bf k}) Im 
G^R_{\alpha\beta}({\bf k},\omega);
\nonumber \\
&& G^R_{\alpha\beta}({\bf k},\omega)=
({\omega-\epsilon_{\bf k}+ \sigma^z g \mu_B B + 
i\tau_0^{-1}})^{-1}_{\alpha\beta}.
\label{eqd}
\end{eqnarray}
The kinetic energy is $\epsilon_{\bf k}=\hbar^2 {\bf k}^2/2m -\epsilon_F$ and
$n_0(\epsilon_{\bf k})$ is the Fermi distribution of electrons.

The total charge $Q_{xx}$ and spin (pointing along the direction of Zeeman field 
or along the $z$-axis)
$M^z_{xx}$ pumped along ${\bf e}_x$ 
direction per 
period are evaluated using Eq.\ref{cscurrent}.
The final results can be expressed in a form similar to Eq.\ref{ctg}.

\begin{eqnarray}
&& Q_{xx}=
\Pi^{lc}
\sum_{\eta,\xi=1,..}^{N} 
\chi_{\eta\xi}
S_{\eta\xi}, M^z_{xx}=
\Pi^{ls}
\sum_{\eta,\xi=1,..}^{N} 
\chi_{\eta\xi} S_{\eta\xi};\nonumber\\ 
&& \Pi^{lc,ls}=
\Pi^{l}(+)\pm\Pi^l(-).
\label{lcp1}
\end{eqnarray}
Alternatively one can obtain the charge transport by directly evaluating Eq.\ref{ctg} taking into account 
Eqs.\ref{asy},\ref{asy1}.

In Eq.\ref{lcp1}, I have introduced two antisymmetric tensors,
\begin{eqnarray}
&& S_{\eta\xi}=-S_{\xi\eta}=\int dg_\eta\wedge dg_\xi=\frac{1}{2}\int_0^{T_0} dT [g_\eta (T)\frac{\partial}{\partial 
T}
g_\xi(T) - g_\xi(T)\frac{\partial}{\partial T} g_\eta(T)]
\nonumber \\
&& \chi_{\eta\xi}=-\chi_{\xi\eta}=
\int\int
\frac{d{\bf r} d{\bf r}'}{L}
V_\eta({\bf r}) \frac{\partial}{\partial x}G({\bf r}, {\bf r}')
V_\xi({\bf r}');
\label{scp}
\end{eqnarray}
And
\begin{equation}
\Pi^{l}(\pm)=\int{d \epsilon_k} D_k\frac{\partial}{\partial 
\epsilon_{\bf k}}
[\frac{1}{D_k}\frac{\partial \nu(\epsilon_{\bf k})}{\partial \epsilon_{\bf 
k}} n_0(\epsilon_{\bf k}\pm g\mu_B B)].
\label{lcp2}
\end{equation}
Here $\nu({\epsilon}_{\bf k})$ is the one-particle density of states
and the volume of structure is $V=L\times L\times L$.
$\Pi^l(\pm)$ depends on the compressibility 
and $\Pi^l(\pm)\chi_{\eta\xi}=\pi_{\eta\xi}$ defines the longitudinal {\em adiabatic 
curvatures} of 
spin-up and spin-down electrons.
Eqs.\ref{lcp1},\ref{scp},\ref{lcp2} are the general results for charge and spin pumping in the 
semiclassical 
limit.

Introducing
$k_{F}$, $\epsilon_{F}$ and $D_0$ as the Fermi momentum, Fermi energy 
and diffusion constant at the Fermi surface respectively,
one rescales all quantities in Eq.\ref{lcp2},

\begin{eqnarray}
&& |{\bf k}|=k_F a, \epsilon_{\bf k}=e(a)\epsilon_F,
D_k=h(a) D_0; \nonumber \\ 
&& \frac{\partial \nu(\epsilon_{\bf k})}{\partial \epsilon_{\bf 
k}}=f(a)\frac{\partial \nu(\epsilon_F)}{\partial \epsilon_F},
n_0(\epsilon_{\bf k})=n_0(e(a)\epsilon_F)
\label{scale}
\end{eqnarray}
where $e(a), h(a)$ and $f(a)$ are intrinsic functions determined by the energy
dispersion. $a$ varies from zero to infinity; at fermi surfaces, $a=1$ and $e(1)=0$ and $h(1)$ $=f(1)=1$.

The longitudinal adiabatic curvatures are

\begin{eqnarray}
&& \Pi^{lc}=
 \frac{\partial  \nu(k_{F})}{\partial
\epsilon_{F}} \pi^{lc},
\Pi^{ls}= \frac{g\mu_B B}{\epsilon_F}
\frac{\partial \nu(k_{F})}{\partial
\epsilon_{F}} \pi^{ls};
\nonumber\\
&& \pi^{lc}=-2\int da h(a)
\frac{\partial 
}{\partial a} [\frac{f(a)}{h(a)}n_0], \;\
 \pi^{ls}=\int da  h(a) \frac{\partial 
}{\partial a} [\frac{f(a)}{h(a)}\frac{1}{\partial_a e(a) }\frac{\partial 
n_0}{\partial a}]. 
\label{scp1}
\end{eqnarray}
I am mainly interested in zero temperature results here and in the following sections.
If $e, f, h$ are assumed to be smooth functions in the vicinity of $a=1$ and their dervatives at $a=1$
are much less than unity, then in the leading order one obtains $\pi^{lc}=2$, and $\pi^{ls}= \partial_a h(1)/\partial_a e(1)$.

Two important general features of spin pumping deserve
some emphases. One is that 
the spin pumping current is zero if 
$g_\eta(t)$ is identical for $\eta=1,2,..N$. More generally one can show
that the pumped charge and spin have to be vanishing if the trajectory of vector
${\bf g}=(g_1,g_2,...g_N)$ in the N-dimensional space encloses
a zero area. 
This is a well known fact emphasized in a few previous occasions
where charge pumping was studied;
the pumping is a pure geometric effect determined by two-form
curvatures and is absent in a one-dimensional parameter space.

The second feature is that the longitudinal
spin current is proportional to the difference between
adiabatic curvatures ($\Pi^l(\pm))$ of spin-up and spin-down 
electrons. Therefore the longitudinal spin pumping efficience
is

\begin{equation}
\epsilon^l=\frac{M^z_{xx}}{Q_{xx}}=\frac{g\mu_B B}{\epsilon_F}\frac{\pi^{ls}}{\pi^{lc}}.
\end{equation}

In the presence of orbital magnetic fields, one can also evaluate
the transverse charge and spin pumping current,

\begin{eqnarray}
&& Q_{yx}=
\Pi^{tc} \sum_{\eta,\xi=1,..}^{N} 
\chi_{\eta\xi}
S_{\eta\xi}, M^z_{yx}=
\Pi^{ts}\sum_{\eta,\xi=1,..}^{N} 
\chi_{\eta\xi} S_{\eta\xi}.\nonumber\\ 
\label{tcp1}
\end{eqnarray}
The transverse adiabatic curvatures which lead to 
these currents are

\begin{eqnarray}
&& \Pi^{tc,ts}=\tau_0\Omega_c
[\Pi^l(+)\pm \Pi^l(-)]. 
\end{eqnarray}
$\Omega_c=e H_{ext}/mc$ is defined as the cyclotron frequency of
external magnetic fields.
Obviously, one can also introduce transverse pumping angle $\theta_C$ and 
transverse spin pumping angle $\theta_S$ in analogy to the usual
Hall angle, 

\begin{eqnarray}
&& \tan \theta_C=\frac{Q_{yx}}{Q_{xx}}=\tau_0\Omega_c,\nonumber \\
&& \tan 
\theta_S=\frac{M^z_{yx}}{Q_{xx}}=\tau_0\Omega_c \frac{g\mu_B 
B}{\epsilon_F}\frac{\pi^{ls}}{\pi^{lc}}.
\end{eqnarray}
Readers can easily confirm these results.
In this scheme, the spin pumping current vanishes in the absence of Zeeman fields
because $\Pi^l({\pm})$ are identical.

\section{Topological spin pumping I: Topological 
Beam splitting}

The key idea of topological spin pumping lies in
the fact that the transverse motion of electrons is not only affected 
by usual orbital magnetic fields, or a gradient in Zeeman fields
but also by spin rotation. 
Compared to Lorentz forces which act on spin-up
and spin-down electrons undiscriminately, the topological 
force induced by spin-rotation does discriminate spin-up
and spin-down electrons as if they are oppositedly "charged".
Indeed as one will see, spin-up and spin-down electrons carry
opposite charges defined with respect to the Pontryagin 
topological fields.
Splitting of spin-up and spin-down electrons
in topological fields therefore is named as topological
Stern-Gerlach splitting ({\em TSGS}).
So before studying the kinetic approach 
to topological spin pumping, let us offer a qualitative picture of
the phenomenon of {\em TSGS}.

Apparently,
spin rotation doesn't occur in free spaces where the electron spin
${\bf S}_z$ is a good quantum number.
So for {\em TSGS} to happen, 
certain mechanism has to be introduced to rotate spins during transport.
It can be achieved by a coupling between electrons 
and an artificial background "magnetic" configuration.
To illustrate this idea of {\em TSGS},
one
studies the following Hamiltonian
\begin{equation}
{\cal H}=\frac{{\bf p}^2}{2m}+V({\bf r})+ g \mu_B B_0 {\bf \sigma}\cdot 
{\bf \Omega}({\bf r}) +V_{ext}({\bf r},T).
\end{equation}
The unit vector ${\bf \Omega}$ is defined by two
angles $\theta({\bf r})$ 
and 
$\phi({\bf r})$ in spherical coordinates;

\begin{eqnarray}
&& {\Omega}_x({\bf r})=\sin\theta(\bf r)\cos\phi({\bf r}),\nonumber\\
&& { \Omega}_y({\bf r})=\sin\theta({\bf r})\sin\phi({\bf r}),\nonumber\\
&& {\Omega}_z({\bf r})=\cos\theta({\bf r}).
\label{unit}
\end{eqnarray}

Consider
the following 
coherent states of electrons 

\begin{eqnarray}
&& |{\bf \Omega}({\bf r});+>=\Psi({\bf r}) \otimes \cos\frac{\theta}{2} 
\exp(-i\frac{\phi}{2}) \exp( 
\tan\frac{\theta({\bf 
r})}{2}e^{i\phi} 
{\bf \sigma}^-) |\uparrow>, \nonumber \\ 
&& |{\bf \Omega}({\bf r});->=\Psi({\bf r})\otimes \sin\frac{\theta}{2} 
\exp(-i\frac{\phi}{2})
\exp(- \cot\frac{\theta({\bf 
r})}{2}e^{i\phi} 
{\bf \sigma}^-) |\uparrow>.
\label{updown}
\end{eqnarray}
$|\uparrow>$ is the spin-up state defined along $z$ axis and $\sigma^-$ is 
the corresponding lowering operator.
$|{\bf \Omega};\pm>$ are spin-up and spin-down states 
defined in a local frame where vector ${\bf e}_z$ coincides with unit 
vector ${\bf 
\Omega}({\bf r})$; i.e.

\begin{equation}
{\bf \Omega}({\bf r}) \cdot \sigma |{\bf \Omega};\pm>=\pm |{\bf \Omega};\pm>
\end{equation}
at every point ${\bf r}$. 
Electrons in these states experience $X$-space spin rotation
({\em XSSR}).

These states are called spin-{\em plus} and spin-{\em minus}
states to be distinguished from spin-up and spin-down states defined 
before.
Obviously, {\em plus} and {\em minus} states discussed above are 
exact eigen states of the local Zeeman coupling and their degeneracy
is lifted at a finite $B_0$.

To demonstrate the beam splitting,   
one evaluates the expectation value of
energy operator ${\cal H}$
in these two sets of states.
The results in this limit can be conveniently cast into
the following form

\begin{equation}
<{\cal H}>_{\pm}=<\Psi({\bf r})|\frac{1}{2m} [{\bf p} \pm
{\bf A}^X({\bf 
r})]^2 +V({\bf r}) |\Psi({\bf r})>\pm \;\ g \mu_B B_0.
\label{Tcharges}
\end{equation}
Here
${\bf p}$ is the momentum operator to be distinguished from the momentum ${\bf k}$ which is a variable.
${\bf A}^X$-fields are confirmed to be the vector 
potential of following  topological fields

\begin{equation}
{\bf T}^X_\lambda=\frac{1}{4}
\epsilon_{\lambda\mu\nu}
{\bf \Omega}({\bf r})\cdot \frac{\partial {\bf \Omega}({\bf 
r})}{\partial {\bf r}_\mu}
\times \frac{\partial {\bf \Omega}({\bf r})}{\partial {\bf r}_\nu}.
\label{Tf0}
\end{equation}
$\epsilon_{\lambda\mu\nu}$ $=-\epsilon_{\mu\lambda\nu}$ 
$=-\epsilon_{\lambda\nu\mu}$ is an antisymmetric 
tensor.
In both Eqs.\ref{Tcharges},\ref{Tf0},I use superscript $X$ to refer to 
$X$-space gauge fields. And in Eq.\ref{Tcharges}, I have neglected 
terms which are identical for $|\pm>$ states.
This form of the kinetic energy of spin-rotating electrons was 
previously derived to demonstrate 
interactions between quasi-particles
and spin fluctuations in triplet superconductors\cite{Zhou02}.

Therefore the total forces exserted on spin-{\em plus} and spin-{\em 
minus} electrons
are
\begin{equation}
{\bf F}_{\pm}=
<\Psi|-\frac{\partial V({\bf r})}{\partial {\bf r}}\pm 
{\bf T}^X\times \frac{{\bf 
p}\pm {\bf A}^X({\bf r})}{m} |\Psi>.
\label{force}
\end{equation}
The last identity holds when 
the spin rotation is adiabatic so that
transitions between Zeeman split spin-{\em plus} and spin-{\em minus} states are 
negligible.
I.E.
\begin{eqnarray}
| \frac{\hbar}{2m} \frac{\partial \phi({\bf r})}{\partial {\bf r}} \cdot 
Im \Psi^*({\bf r}) \frac{\partial}{\partial {\bf r}} \Psi({\bf r})|
\ll g \mu_B B_0.
\label{ad1}
\end{eqnarray}
One can verify that when this adiabaticity condition is satisfied,
$\sigma\cdot {\bf n}$ is an approximate {\em good quantum number}.

In the semiclassical approximation employed below, I further assume
that

\begin{equation}
|Im \Psi^*({\bf r}) \frac{\partial}{\partial {\bf r}} \Psi({\bf r})| \gg 
|\frac{\partial
\theta({\bf r})}{\partial {\bf r}}|, 
|\frac{\partial \phi({\bf r})
}{\partial {\bf r}}|
\end{equation}
So over the wave length of electrons, spin rotation is negligible,

In addition to a term proportional to the field gradient of scalar fields, there is a
new force perpendicular to 
the velocity of electrons
similar to the Lorentz force. More important
as indicated in Eq.\ref{Tcharges}, spin-{\em plus} and spin-{\em minus} 
electrons carry
opposite topological charges; therefore, the corresponding forces are 
in fact along exactly opposite directions.
This shows that an {\em XSSR} does affect the orbital
motion of electrons and does differentiate spin-{\em plus} states from
spin-{\em minus} states. It therefore leads to the 
promised phenomenon of {\em TSGS}.

It is important to further emphasize here that 
to observe {\em TSGS}, the background configuration has to be
topological nontrivial (see more in the next section)
so that ${\bf T}^X$ is nonzero.
To highlight the relevance of topology of spin configurations,
let us consider spin states defined in Eq.\ref{updown}
where ${\bf \Omega}({\bf r})$
corresponds to a hedgehog configuration,

\begin{equation}
{\bf \Omega}(\bf r)=\frac{\bf r}{|{\bf r}|}.
\end{equation}
Our calculations show that
forces acting on two spin-rotating electrons
given in Eq.\ref{updown} are equivalent to forces exserted on 
two {\em oppositely charged} particles in
a resultant magnetic monopole field

\begin{equation}
{\bf T}^X=\frac{\bf r}{2|{\bf r}|^3}.
\end{equation}

Before leaving this section,
I generalize the argument to momentum space topological effects.
I then consider two orthogonal ${\bf k}$-space wave packets;
spins in these states are pointing at either ${\bf k}$ or $-{\bf k}$
direction.
Let us define $\theta({\bf k})$ and $\phi({\bf k})$ as
\begin{eqnarray}
&& {\bf \Omega}_x({\bf k})=\sin\theta(\bf k)\cos\phi({\bf k}),
\nonumber\\
&& {\bf \Omega}_y({\bf k})=\sin\theta({\bf k})\sin\phi({\bf k}),
\nonumber\\ && {\bf \Omega}_z({\bf k})=\cos\theta({\bf k}).
\end{eqnarray}
and ${\bf \Omega}({\bf k})={\bf k}/|{\bf k}|$ is a unit vector 
along ${\bf k}$.

\begin{eqnarray}
&& |{\bf \Omega}({\bf k});+>=\Psi({\bf k}) \cos\frac{\theta}{2}
\exp(-i\frac{\phi}{2}) \exp( 
\tan\frac{\theta({\bf 
k})}{2}e^{i\phi} 
{\bf \sigma}^-) |\uparrow>, \nonumber \\ 
&& |{\bf \Omega}({\bf k});->=\Psi({\bf k})
\exp(-i\frac{\phi}{2})
\sin\frac{\theta}{2}
\exp(-\cot\frac{\theta({\bf 
k})}{2}e^{i\phi} 
{\bf \sigma}^-) |\uparrow>.
\end{eqnarray}
As before I assume these spin-{\em plus} and spin-{\em minus} states
are split by an effective Zeeman field $B_0$.

An electron in such a wave packet experiences ${\bf k}$-space spin 
rotation ({\em KSSR}), or spin rotation that depends on its momentum.
For the same reason mentioned before, I further assume adiabaticity in
spin rotation and use a semiclassical approximation. This requires that
\begin{eqnarray}
&&  g\mu_B B_0 \gg
| \frac{\partial }{\partial {\bf k}} \phi({\bf k}) 
\cdot \frac{\partial}{\partial {\bf r}} H({\bf r})|,
\nonumber\\
&& | Im \Psi^*({\bf k}) \frac{\partial}{\partial {\bf k}} \Psi({\bf k})| 
\gg 
|\frac{\partial
\theta({\bf k})}{\partial {\bf k}} 
|, |\frac{\partial
\phi({\bf k})}{\partial {\bf k}}|. 
\nonumber\\
\label{ad2}
\end{eqnarray}

The group velocities of these wave packets are

\begin{eqnarray}
{\bf v}_{\pm}=i<[{\bf r},
{\cal H}({\bf r},{\bf p})]_-
>_{\pm}=<\Psi({\bf 
k})|\frac{{\bf k}}{m}
\mp [\nabla \times {\bf A}^K({\bf k})]\times 
\frac{\partial H({\bf k};{\bf r})}{\partial {\bf r}}|\Psi({\bf k})>.
\label{velocity}
\end{eqnarray}
Superscript $K$ is introduced to specify the ${\bf k}$-space gauge fields.
So the velocity does acquire an additional nontrivial transverse term
in the presence of an external field gradient and topologically
nontrivial fields ${\bf T}^K({\bf k})$ of which ${\bf A}^K({\bf k})$ is the vector potential. A calculation shows that
\begin{equation}
{\bf T}^K_{\lambda}=\frac{1}{4}\epsilon_{\lambda\mu\nu}
{\bf \Omega}({\bf k})\cdot \frac{\partial {\bf \Omega}({\bf k})}{\partial {\bf k}_\mu}
\times \frac{\partial {\bf \Omega}({\bf k})}{\partial {\bf k}_\nu}.
\label{pyf}
\end{equation}
And more important the spin-up and spin-down electrons defined in the local frames
drift in an opposite direction once again leading to
{\em TSGS}.

The general property of electrons illustrated in Eq.\ref{velocity} has been noticed in a few
different occasions. In Ref. \cite{Sundaram99,Jungwirth02},
${\bf T}^K$ fields were expressed as Chern-number density of 
electron states which was first introduced for the studies of quantum Hall and fractional 
quantum quantum Hall conductances \cite{Thouless82}.  
In Ref.\cite{Murakami03}, ${\bf T}^K$ fields are from a 
topological monopole
in the ${\bf k}$-space.

The general form of topological fields obtained in Eq.\ref{pyf} is
a generalization of standard Pontryagin fields defined in the $X$-space.
One easily recognizes that topological fields discussed here are equivalent 
to Berry's two-form fields defined in an external parameter 
space\cite{Berry84}. They 
generally represent holonomy of parallelly transporting  
eigen vectors in the Hilbert space\cite{Simon83}, in this particular
case the holonomy of transporting a spin-up or spin-down eigenvector defined in
local frames. I will discuss topological spin pumping in the
presence of this {\em TSGS} in section VI.

\section{Topological spin pumping II:
adiabatic spin transport in the presence of {\em XSSR}}

I now take into the kinetics which leads to spin
rotation while electron wave packets propagate
in the $X$-space. I study the spin pumping phenomena 
in this case via applying kinetic equations 
derived in section II.

Consider electrons coupled to a background spin configuration
or an artificial magnetic field with uniform magnitude but with spatially 
varying orientation.
One models electrons with the following Hamiltonian

\begin{eqnarray}
&& H({\bf k},{\bf r},T)=\frac{{\bf k}^2}{2m}-\mu_F -
g\mu_B  [B_0{\bf \Omega}({\bf r}) \cdot {\bf \sigma}+
B \sigma_z ] +V_{ext}({\bf r},T).
\end{eqnarray}
again ${\bf \Omega}({\bf r})$ is a unit vector representing the 
orientation of
an internal exchange field in metals.
The scattering of impurity potentials is taken into account in elastic
collision integrals in Eq.4.
I assume $g\mu_B B_0, g\mu_B B$ are much smaller 
than the fermi energy $\epsilon_F$.

To facilitate discussions, one introduces the following unit vector
${\bf n}({\bf r})$ which defines the direction of net Zeeman fields in the above equation

\begin{eqnarray}
&& {\bf n}_x=\frac{{\Omega}_x}
{\sqrt{{\Omega}_x^2+{\Omega}_y^2+(\Omega_z + I)^2}}
\nonumber  \\
&& {\bf n}_y=\frac{{\Omega}_y}{\sqrt{{\Omega}_x^2+{\Omega}_y^2+(\Omega_z + I)^2}}
\nonumber \\
&& {\bf n}_z=\frac{{ \Omega}_z +I}{\sqrt{{ \Omega}_x^2+{ \Omega}_y^2+(\Omega_z + I)^2}}
\label{n1}
\end{eqnarray}
where $I=B/B_0$.

Alternatively,
similar to Eq.\ref{unit}, in spherical coordinates one introduces the following 
characterization of
${\bf n}$, 
\begin{eqnarray}
&& {\bf n}_x=\sin\tilde{\theta}({\bf r})\cos\tilde{\phi}({\bf 
r}),\nonumber\\
&&
{\bf n}_y=\sin\tilde{\theta}({\bf r})\sin\tilde{\phi}({\bf r}),\nonumber\\
&& {\bf n}_z=\cos\tilde{\theta}({\bf r}).
\end{eqnarray}
I use superscript {\em tilde} to distinguish the spherical coordinates
$\tilde{\theta}$,$\tilde{\phi}$ for ${\bf n}$
from $\theta$,$\phi$ for ${\bf \Omega}$.
Eq.\ref{n1} indicates that

\begin{equation}
\tilde{\phi}=\phi({\bf r}),\tilde{\theta}=\arctan \frac{\sin\theta({\bf r})}{\cos\theta({\bf r}) +I}.
\end{equation}
Introduce a local spin rotation such that 
\begin{eqnarray}
U^{-1}({\bf r})({\bf n}({\bf r})\cdot {\bf \sigma})U({\bf r})=
{\bf \sigma}^z.
\end{eqnarray}
The Hamiltonian becomes

\begin{eqnarray}
H({\bf k},{\bf r},T)=\frac{ ({\bf k}-{\bf A}^X({\bf 
r}))^2}{2m}-\mu_F-g \mu_B B_0 [{\Omega}_x^2+{\Omega}_y^2+(\Omega_z + 
I)^2]^{1/2}\sigma_z +V_{ext}({\bf r}, T).
\label{rH2}
\end{eqnarray}
Here $SU(2)$ gauge fields generated by pure spin rotation
are 

\begin{eqnarray}
{\bf A}^X_{\mu}=iU^{-1}(\bf 
r)\frac{\partial}{\partial {\bf 
r}_\mu} U({\bf r})
={\bf A}^{X\gamma}_{\mu}  {\bf \sigma}^{\gamma}.
\end{eqnarray}
$\gamma=x,y,z$.
To simplify the formula,
in this equation and the rest of article I do not show spin indices explicitly.
A direct calculation yields

\begin{eqnarray}
&& {\bf A}^{Xx}_\mu=-\frac{1}{2}{\sin\tilde{\theta}({\bf 
r})}\frac{\partial \tilde{\phi}({\bf r})}{\partial r_\mu},\nonumber\\
&& {\bf A}^{Xy}_\mu=\frac{1}{2}
\frac{\partial \tilde{\theta}({\bf r})}{\partial r_\mu},\nonumber\\
&& {\bf A}^{Xz}_\mu=\frac{1}{2}{\cos\tilde{\theta}({\bf 
r})}\frac{\partial \tilde{\phi}({\bf r})}{\partial r_\mu};
\end{eqnarray}
And {\em full} covariant SU(2) fields $\tilde{\Sigma}^X_{\mu\nu}$ 
vanish as one should expect for pure spin rotation.

To proceed further, one notes that 
the degeneracy between spin-up and spin-down states in the rotated basis
or spin-{\em plus} and spin-{\em minus} states is 
completely lifted by various Zeeman fields.
One again assumes that spin rotation is slow in the $X$-space
so that 
the adiabaticity specified in Eq.\ref{ad1} can be satisfied.
In the adiabatic approximation, one neglects 
transitions between spin-{\em plus} and spin-{\em minus} states
and set off-diagonal gauge potentials ${\bf A}^{Xx}$,${\bf 
A}^{Xy}$ as zero. 
Therefore in Eq.\ref{rH2}, one only keeps ${\bf A}^{Xz}$ which 
yields the usual Berry curvatures for spin-{\em plus} and spin-{\em minus} states.
Corresponding $U(1)$ gauge fields are

\begin{eqnarray}
&& {\Sigma}^{Xz}_{\mu\nu}(\bf r)=\frac{\partial {\bf 
A}^{Xz}_{\mu}({\bf 
r})}{\partial 
{\bf r}_\nu}-\frac{\partial {\bf A}^{Xz}_{\nu}({\bf r})}{\partial 
{\bf r}_\mu} 
%+\frac{1}{i}[{\bf A}^X_\mu({\bf r}), {\bf A}^X_\nu({\bf 
%r})]_{-}.
\label{naf}
\end{eqnarray}

So in this limit, only the $z$-component of SU(2) gauge potentials survives 
to contribute to
pumping currents; it also
defines well known Pontryagin type $U(1)$ fields 

\begin{eqnarray}
&& {\bf \Sigma}^X_{\mu\nu}={\bf \Sigma}_{\mu\nu}^{Xz}\sigma^z,
\;\
{\bf \Sigma}^{Xz}_{\mu\nu}
=\frac{1}{2}\sin\tilde{\theta} ({\bf r})
[\frac{\partial \tilde{\theta}({\bf r})}{\partial {\bf 
r}_\mu}\frac{\partial \tilde{\phi}({\bf r})}{\partial {\bf r}_\nu}
-\frac{\partial \tilde{\theta}({\bf r})}{\partial {\bf 
r}_\nu}\frac{\partial\tilde{\phi}({\bf r})}{\partial {\bf r}_\mu}].
\label{pyf1}
\end{eqnarray}
%which is equivelant to
%\begin{equation}
%{\bf \Sigma}^{Xz}_{\mu\nu}=
%\frac{1}{2}{\bf n}\cdot \frac{\partial {\bf n}}{\partial {\bf X}_\mu}
%\times \frac{\partial {\bf n}}{\partial {\bf X}_\nu}.
%\end{equation}
$\tilde{\theta}$ and $\tilde{\phi}$ again are two spherical angles
of ${\bf n}({\bf r})$ in Eq.46.

In the rotated basis, the structure of equation for 
one-particle density matrix should be identical to 
the one
in the previous section. 
However according to a general consideration in the
semiclassical transport theory, the following 
transformation takes place in the equation of 
motion for electrons in the presence of {\em XSSR}.

\begin{eqnarray}
&& -\frac{\partial H({\bf k},{\bf r};T)}{\partial {\bf r}}
\rightarrow 
\frac{\partial {\bf k}}{\partial t}=
[{\bf k},H({\bf k},
{\bf r};T)]_{{\bf r},{\bf p}}; \nonumber \\
&& \frac{\partial H({\bf k},{\bf r};T)}{\partial {\bf k}}
\rightarrow 
\frac{\partial {\bf r}}{\partial t}=
[{\bf r}, H({\bf k}, {\bf r};T)]_{{\bf r},{\bf p}}.
\end{eqnarray}
$[]_{u,v}$ is the usual Poisson bracket defined with respect to canonical 
coordinates $\{u,v \}$.
${\bf k}$ is the electron momentum 

\begin{equation}
{\bf k}={\bf p} +{\bf A}^{Xz}({\bf r})\sigma_z; 
\end{equation}
and ${\bf p}$ and ${\bf r}$ are a pair of canonical coordinates. 
In the semiclassical approximation, the new equation for the one-particle density matrix therefore looks as

\begin{eqnarray}
&&\frac{\partial \rho({\bf k},{\bf r};\omega,T)}{\partial 
T}- \{
[\frac{\partial H({\bf k},{\bf r};T)}{\partial
{\bf r}_\mu}+ 
\Sigma^X_{\mu\nu}({\bf r}) \otimes
\frac{\partial H({\bf k},{\bf 
r};T)}{\partial {\bf k}_\nu} ]
\frac{\partial}{\partial {\bf k}_\mu}
\nonumber \\
&& \frac{\partial H({\bf k},{\bf r};T)}{\partial {\bf 
k}_\mu}
\frac{\partial}{\partial {\bf r}_\mu}
+\frac{\partial H({\bf k},{\bf r};T)}{\partial T}
\frac{\partial}{\partial \omega} \}\otimes
\rho({\bf k},{\bf r};\omega,T) ={\cal I}^{C.I.}\rho({\bf k},{\bf r};\omega,T)
\end{eqnarray}

Following Appendix A,
the transverse pumping currents are given as

\begin{eqnarray}
&& {\bf J}_\mu({\bf r})=\frac{ \tau_0 }{m}\int {d\omega} 
\frac{d^d{\bf 
k}}{(2\pi)^d} D_k
tr \{ \Sigma^{X}_{\mu\nu}\otimes
\frac{\partial 
H({\bf k}, {\bf r};T)
}{\partial {\bf r}_\nu}
\frac{\partial}{\partial \epsilon_k}[\frac{1}{D_k}
M^{S1}({\bf           
k},{\bf r};\omega,T)] \}
%[ G({\bf X},{\bf X}')
%\frac{\partial H({\bf k},{\bf X};T)}{\partial T}
%\otimes
%\frac{\partial^2 \rho^0({\bf k},{\bf 
%X};\omega,T)}{\partial^2 \omega \partial \epsilon_k}] 
\nonumber \\
&& {\bf J}^z_\mu({\bf r})=\frac{\tau_0 }{2 m}\int d\omega 
\frac{d^d{\bf 
k}}{(2\pi)^d} D_k tr \{ {\bf n}({\bf r}) \cdot {\bf \sigma}
\Sigma^{X}_{\mu\nu} \otimes
 \frac{\partial 
H({\bf k}, {\bf r};T)
}{\partial {\bf r}_\nu}
\frac{\partial}{\partial \epsilon_k}
[\frac{1}{D_k}M^{S1}({\bf           
k},{\bf r};\omega,T)] \}.
%[G({\bf X},{\bf X}')
%\frac{\partial H({\bf k},{\bf X};T)}{\partial T}
%\otimes \frac{\partial^2 \rho^0({\bf k},{\bf 
%X};\omega,T)}{\partial^2 \omega \partial \epsilon_k}] \nonumber \\
\label{TMX}
\end{eqnarray}
where $tr$ is only taken over the spin space.
This is the central result 
for topological transverse spin and charge pumping in the presence of {\em 
XSSR}.

Let me now again consider external perturbations specified in Eq.13.
To address spin pumping, I consider a background spin configuration of square 
half-skyrmion 
lattice given 
below,

\begin{equation}
{\bf n}_x ({\bf r}) +i {\bf n}_y({\bf r})
=\prod_{l_1,l_2}\frac{z
-z(l_1,l_2)}{\sqrt{|z-z(l_1,l_2)|^2+\lambda_S^2}};
z={\bf r}_x+i {\bf r}_y. 
\end{equation}
$z$ is introduced
as a coordinate in the
2D plane. $z(l_1,l_2)=l_1 a  +i l_2 a$ represents a lattice site
with $l_{1,2}$ as integers and $a$ the lattice constant. I also assume that $\lambda_S \ll a$.
The average topological fields of this lattice are 

\begin{eqnarray}
\Sigma_{xy}^{Xz}=\Sigma_0=\frac{\pi}{a^2}.
\label{TF}
\end{eqnarray}
The adiabaticity condition in Eq.\ref{ad1} requires that

\begin{equation}
|{\bf k}| \frac{\hbar}{ m \lambda_s} 
\ll g\mu_B min\{ B_0, B\}.
\end{equation}

The equilibrium density matrix is still given in Eq.\ref{eqd}.
In the longitudinal direction the results are the same as in Eq.24 and I 
will not repeat here. Furthermore,
some straightforward calculations lead to
the following expression for total charge and spin transported in
a transverse direction $y$ per period

\begin{eqnarray}
&&Q_{yx}=
\Pi^{tc}\sum_{\eta,\xi=1,..}^{N} 
\chi_{\eta\xi}
S_{\eta\xi}, M^z_{yx}= 
\Pi^{ts}\sum_{\eta,\xi=1,..}^{N} 
\chi_{\eta\xi} S_{\eta\xi};\nonumber \\
&& \Pi^{tc,ts}=\tau_0 \Omega_{ct} 
[\Pi^{l}(+)
\mp\Pi^l(-)],
\Omega_{ct}=\frac{\Sigma_0}{m}.
\end{eqnarray}
And $\Omega_{ct}$ is introduced as an effective cyclotron frequency of
topological fields $\Sigma_0$.

The transverse charge pumping angle $\theta_C$ and 
transverse spin pumping angle $\theta_S$ 
are 

\begin{eqnarray}
&& \tan \theta_C=\tau_0\Omega_{ct} \frac{g \mu_B B}{\epsilon_F} 
F(\frac{B_0}{B}), \nonumber \\
&&\tan \theta_S=\tau_0\Omega_{ct}.
\end{eqnarray}
$F(\beta)$ is a function of $\beta$; it approaches unity as $\beta$ becomes
much less than one.
The transversal spin pumping efficiency in this case is

\begin{equation}
\epsilon^t=\frac{M^z_{yx}}{Q_{yx}}=\frac{\tan \theta_S}{\tan \theta_C}=
\frac{\epsilon_F}{g\mu_B B} F^{-1}(\frac{B_0}{B}).
\end{equation}

It is important to notice that $\epsilon^t$ diverges as the
external Zeeman field $B$ goes to zero signifying zero transverse charge pumping.
In practical cases, the topological spin pumping  
is always accompanied by small transverse charge pumping current,
a unique feature of topological fields. 
This also occurs in the scheme discussed in the next section.
It is however 
in contrast to spin pumping of polarized electrons
discussed in section II where the transverse spin pumping current is
negligible compared to the transversal charge pumping current.
This feature originates from the fact that spin-{\em plus} and 
spin-{\em minus}
electrons carry opposite topological charges. In the absence of 
polarization, there are equal amplitude of currents 
of spin-{\em plus} and spin 
{\em minus} electron flow in opposite transverse directions as a result of 
{\em TSGS}. So the total
charge pumping is zero but spin pumping current flows.

In all topological pumps discussed here and below, I have found that spins are pumped out
by applied {\em a.c.}gate voltages because of 
a spin configuration which yields either nonzero $\Sigma^X_{\mu\nu}$ 
as in Eq.\ref{TMX} 
or nonzero $\Sigma^K_{\mu\nu}$ as in Eq.\ref{TMK}.
I intend to call these topological configurations
as {\em topological motors} in spin pumps.

\section{Topological spin pumping III: 
adiabatic spin transport in the presence of {\em KSSR}}

To illustrate the effect of {\em KSSR}, I start with 
a 3D toy Model and discuss the topological mechanism.
In the second half of this section, I study the topological mechanism
in more realistic models for electrons in semiconductors.

In all subsections here I assume that spin-orbit splitting and Zeeman field splitting
are much smaller than the fermi energy but can be comparable between themselves.
Zeeman splitting is introduced to ensure that spin degeneracies at ${\bf k}=0$ are lifted and
the adiabaticity holds for every state below fermi surfaces.
The zero field limit should be taken when the adiabaticity conditions in Eq.\ref{ad4} are
satisfied.

In the presence of impurity scattering, to ensure adiabaticity I always assume that the impurity potentials are weak compared
with the splitting between two spin bands either due to spin-orbit coupling or due to Zeeman field splitting.
I neglect therefore the interband transitions due to nonadiabatic corrections. Because of this reason, I only consider the limit of strong spin-orbit coupling
and present results at zero temperature.

\subsection{A toy Model}

In this section I consider electron spins coupled 
to the momenta of electrons and spin rotation
occurs when a wave packet propagates in the momentum space.
A propagating wave packet in the ${\bf k}$-space corresponds to an accelerated 
electron.
The artificial model I introduce here to study topological 
spin pumping can be considered as a 
mathematical generalization
of the Luttinger Hamiltonian\cite{Luttinger56} to spin-1/2 electrons.

Consider the Hamiltonian

\begin{eqnarray}
&& H({\bf k},{\bf r};T)=
\epsilon_{\bf k}- g \mu_B B_0 (|{\bf k}|)
[I (|{\bf k}|) \sigma_z+
{\bf \Omega}({\bf k}) \cdot {\bf \sigma}]
+V_{ext}({\bf r};T);
\label{toy}
\end{eqnarray}
$I$ and $B_0$ are functions of $|{\bf k}|$
and ${\Omega}({\bf k})$ is a unit vector along the direction of ${\bf k}$;

\begin{equation}
I=\frac{B}{B_0 (|{\bf k}|)}, \;\ B_0(|{\bf k}|)=\Gamma |{\bf k}|,\;\
{\bf \Omega}({\bf k})=\frac{\bf k}{|{\bf k}|}.
\label{I}
\end{equation}
In the toy model there are two spin dependent terms;
the term proportional to $\sigma_z$ represents the Zeeman splitting of 
electrons in the presence of
fields along the $z$-axis and ${\bf \Omega} \cdot {\bf \sigma}$ term 
characterizes a collinear spin-orbit correlation. 

To understand the topology of spin configurations in Fermi seas, I again introduce
a unit vector ${\bf n}({\bf k})$ as a function of ${\bf \Omega}({\bf k})$ 
defined in Eq.\ref{n1} in the previous 
section.
Especially,

\begin{eqnarray}                                                               
&& \tilde{\phi}({\bf k})=\phi({\bf k}),
\nonumber \\
&& \tilde{\theta}=\arctan 
\frac{\sin\theta({\bf k})}{\cos\theta({\bf k}) +I(|{\bf k}|)}.     
\end{eqnarray}                                                                 
Again $\theta({\bf k}),\phi({\bf k})$ are two spherical coordinates          
of unit vect $\Omega({\bf k})={\bf k}/|{\bf k}|$ and 
differ from two angles of ${\bf n}({\bf k})$, $\tilde{\theta}({\bf 
k})$ and $\tilde{\phi}({\bf k})$ (see Eq.\ref{unit}).

One then considers a configuration of ${\bf n}({\bf k})$ on a sphere 
at a very large momentum which 
naturally
defines a mapping from an external $S^2$ sphere in the ${\bf k}$-space to 
a target $S^2$ space where ${\bf n}({\bf k})$
lives. 
The topology of electron spin states in Fermi seas is therefore 
characterized by 
$\pi_2(S^2)$,
the second group of target space $S^2$.

Let us further introduce ${\bf T}^K$ fields defined in Eq.\ref{pyf} with ${\bf \Omega}({\bf k})$ replaced with 
${\bf n}({\bf k})$. 
The winding number of a mapping or configuration can be
characterized by 
the flux of ${\bf T}^K$ fields 
through a large surface. It is easy to verify
that

\begin{equation}
\frac{1}{2\pi}\oint_S dS \frac{{\bf k}}{|{\bf k}|}\cdot {\bf T}^K=1
\end{equation}
where the surface integral is taken over at $|{\bf k}|\rightarrow +\infty$. 
This shows that spins of electrons form a monopole structure.
This is not surprising because at very large ${\bf k}$ or small $I$, ${\bf n}({\bf k})$ is identical
to ${\bf \Omega}({\bf k})$ which always points outward along the radius direction.

Again introduce a ${\bf k}$-space local spin rotation
such that 
\begin{eqnarray}
U^{-1}({\bf k}){\bf n}({\bf k}) \cdot {\bf \sigma} U({\bf k})=
{\bf \sigma}^z.
\label{ksr}
\end{eqnarray}
Under the spin rotation,
the Hamiltonian becomes

\begin{eqnarray}
H({\bf k},{\bf r};T)=\epsilon_{\bf k}
-g \mu_B B_0 [{\Omega}_x^2+{\Omega}_y^2+(\Omega_z + I(\rho_{\bf k}))^2]^{1/2}\sigma_z
+V_{ext}({\bf r}-{\bf A}^K;T).
\label{rH1}
\end{eqnarray}

As before,
SU(2) gauge fields are generated under the spin rotation

\begin{eqnarray}
&& {\bf A}^K_{\mu}({\bf k})
=iU^{-1}(\bf k)\frac{\partial}{\partial {\bf k}_\mu} 
U({\bf k})
={\bf A}^{K\gamma}_{\mu} {\bf \sigma}^{\gamma};
\end{eqnarray}
in a fixed gauge, one obtains

\begin{eqnarray}
&& {\bf A}^{Kx}_\mu=-\frac{1}{2}{\sin\tilde{\theta}({\bf
k})}\frac{\partial \tilde{\phi}({\bf k})}{\partial {\bf k}_\mu},
\nonumber\\
&&{\bf A}^{Ky}_\mu=\frac{1}{2}
\frac{\partial \tilde{\theta}({\bf k})}{\partial {\bf k}_\mu},
\nonumber\\
&& {\bf A}^{Kz}_\mu=\frac{1}{2}{\cos\tilde{\theta}({\bf
k})}\frac{\partial \tilde{\phi}({\bf k})}{\partial {\bf k}_\mu}.
\label{fgf}
\end{eqnarray}                         
And the {\em full} SU(2) fields again vanish.

In linear responses, the effective Zeeman field splitting between 
spin-{\em plus} and spin-{\em minus} states in Eq.\ref{rH1} is 
stronger than external perturbations.
Further, I require that the energy splitting is also stronger than impurity potentials.
So the adiabaticity condition in
Eq.\ref{ad2} is always satisfied. I therefore set ${\bf A}^{Kx}$,
${\bf A}^{ky}$ to be zero again to neglect transitions between spin-{\em 
plus}
and spin-{\em minus} states. In Eq.\ref{rH1} I only keep the
$z$-component of SU(2) potentials ${\bf A}^{Kz}_{\mu\nu}$ which yields to 
Berry's phases of {\em plus} and {\em minus} states.
The corresponding $z$-component of reduced SU(2) gauge potentials which enters 
our results below is

\begin{eqnarray}
&& {\Sigma}^{Kz}_{\mu\nu}(\bf k)=\frac{\partial {\bf 
A}^{Kz}_{\mu}({\bf 
k})}{\partial 
{\bf k}_\nu}-\frac{\partial {\bf A}^{Kz}_{\nu}({\bf k})}{\partial 
{\bf k}_\mu}.
%+\frac{1}{i}[{\bf A}^K_{\mu}({\bf k}), {\bf A}^K_{\nu}({\bf 
%k})].
\end{eqnarray}

I note that in the presence of spin-orbit coupling, impurity scattering
combined with ${\bf A}^{Kx,Ky}$ components of $SU(2)$ gauge fields does lead to transitions between different spin bands. 
The adiabaticity condition in this case however is sufficient to ensure that these contributions are negligible.
So the Born-approximation employed in this article is valid in the strong spin-orbit coupling and finite Zeeman field limit where the splitting between ${\em plus}$ 
and ${\em minus}$ spin bands at any momentum ${\bf k}$,

\begin{equation}
2 g \mu_B \Gamma |{\bf k}| [{\Omega}_x^2+{\Omega}_y^2+(\Omega_z + I(\rho_{\bf k}))^2]^{1/2}
\end{equation}
is much stronger than the impurity potentials.

Where the adiabaticity condition is satisfied,
one shows that

\begin{eqnarray}
&& {\bf \Sigma}^{K}_{\mu\nu}=
{\bf \Sigma}^{Kz}_{\mu\nu}\sigma_z, \;\
{\bf \Sigma}^{Kz}_{\mu\nu}=
\frac{1}{2}{\bf n}({\bf k}) \cdot \frac{\partial {\bf n({\bf k})}}{\partial {\bf k}_\mu}
\times \frac{\partial {\bf n}({\bf k})}{\partial {\bf k}_\nu}.
\nonumber \\
\label{monopole}
\end{eqnarray}
In spherical coordinates where ${\bf k}=(\rho_{\bf k},\theta,\phi)$, one has the following explicit results

\begin{eqnarray}
&& \Sigma^{Kz}_{\theta\phi}=
%\frac{1}{2 \rho_{\bf 
%k}^2}\frac{\sin\tilde{\theta}}{\sin{\theta}}\frac{\partial 
%\tilde{\theta}}{\partial \theta}=
 \frac{1+I \cos\theta}{(1+I^2 +2 \cos\theta I)^{3/2}}
\frac{1}{2\rho^2_{\bf k}}
\nonumber \\
&& \Sigma^{Kz}_{\rho\phi}=
%\frac{\sin\tilde{\theta}}{2\rho_{\bf k}\sin\theta}\frac{\partial 
%\tilde{\theta}}{\partial 
%{\rho}_{\bf k}}=
-\frac{\sin\theta}{(1+I^2 +2 \cos\theta I)^{3/2}}
\frac{1}{2\rho_{\bf k}} \frac{\partial I({\rho}_{\bf k})}
{\partial \rho_{\bf k}}
\label{monopole1}
\end{eqnarray}
and $I$ is a function of $\rho_{\bf k}$ ($=|{\bf k}|$) as defined in
Eq.\ref{I}.

It is convenient to redefine topological fields in terms of ${\bf T}^K_{\mu}$,

\begin{equation}
{\bf T}^K_{\mu}({\bf k})=\frac{1}{2}\epsilon_{\mu\nu\lambda}\Sigma^{Kz}_{\nu\lambda}({\bf k}).
\end{equation}
One finds the following asymptotic behaviors at large and small momenta,

\begin{equation}
{\bf T}^K({\bf k})=\left\{\begin{array}{cc}
\frac{1}{2 \rho^2_{\bf k}}[1-\frac{2\lambda_m}{\rho_{\bf k}}\cos\theta]
{\bf e}_\rho  - \frac{\lambda_m}{2 \rho^3_{\bf k}} \sin\theta {\bf e}_\theta,
&  \mbox{when $\rho_{\bf k} \gg \lambda_m$;} \nonumber\\
\frac{1}{2 \lambda_m^2}{\bf e}_z +\frac{\rho_k}{2 \lambda_m^3}
[(1-3cos^2\theta){\bf e}_\rho + 3\sin\theta\cos\theta {\bf e}_\theta],
&  \mbox{when $\rho_{\bf k} \ll \lambda_m$.}\\
\end{array} \right .
\end{equation}
And here

\begin{eqnarray}
\lambda_m=\frac{B}{\Gamma}
\end{eqnarray}
defines the core of 
anisotropic monopoles when the Zeeman fields are present.

When the Zeeman field $B$ is set to zero or $I=0$, unit vector ${\bf n}({\bf k})$ 
coincides with $\Omega({\bf k})$ and 
$\tilde{\phi}=\phi({\bf k})$, ${\tilde{\theta}}=\theta({\bf k})$; 
therefore, $\Sigma^{Kz}_{\rho\phi}$ vanishes.
Eq.\ref{monopole1} in this limit
indicates familiar isotropic monopole fields in the ${\bf k}$-space.
In the presence of finite Zeeman fields, topological fields are 
not strictly isotropic
because the inversion symmetry $z \rightarrow -z$ is broken by external 
Zeeman fields. Topological fields are along the $z$-axis at small 
momentum limit but approach isotropic monopole fields at large momenta.
The crossover takes place at $\lambda_{m}$.

Under {\em KSSR}, the following transformation should 
occur in the equation for the density matrix,

\begin{eqnarray}
&& - \frac{\partial H({\bf k},{\bf r};T)}{\partial {\bf r}}
\rightarrow 
\frac{\partial {\bf k}}{\partial t}=
[{\bf k},{H}({\bf k}, {\bf r};T)]_{{\bf x},{\bf k}};\nonumber 
\\
&& \frac{\partial H({\bf k},{\bf r};T)}{\partial {\bf k}}
\rightarrow 
\frac{\partial {\bf r}}{\partial t}=
[{\bf r},H({\bf k}, {\bf r}; T)]_{{\bf x},{\bf k}}.
\end{eqnarray}
The electron coordinate in the presence of {\em KSSR} is

\begin{equation}
{\bf r}={\bf x} -{\bf A}^{Kz}({\bf k})\sigma_z; 
\end{equation}
${\bf x},{\bf k}$ are a pair of canonical coordinates in this case.

The corresponding kinetic equation becomes

\begin{eqnarray}
&&\frac{\partial \rho({\bf k},{\bf r};\omega,T)}{\partial T}+ [v_k {\bf \Omega}_{\mu}({\bf k})     
+ \Sigma^K_{\mu\nu}({\bf k})
\otimes \frac{\partial H({\bf k},{\bf r};T)}{\partial 
{\bf       
r}_\nu}] 
\frac{\partial}{\partial {\bf r}_\mu}\nonumber \\
&&-\frac{\partial H({\bf k},{\bf r};T)}{\partial {\bf r}_\mu}
\frac{\partial}{\partial {\bf k}_\mu}
+ \frac{\partial H({\bf k},{\bf r};T)}{\partial T}
\frac{\partial}{\partial \omega} \}\otimes
\rho({\bf k},{\bf r};\omega,T)={\cal I}^{C.I}\rho({\bf k},{\bf r};\omega,T)
\end{eqnarray}

The charge and spin current expressions transform accordingly;
in the rotated basis one has
\begin{eqnarray}
&& {\bf J}({\bf r}, T)=
\int \frac{d\omega}{2\pi} \frac{d^3{\bf k}}{(2\pi)^3}
Tr \{ [v_k {\bf \Omega}_{\mu}({\bf k}) 
+\Sigma^K_{\mu\nu}({\bf k}) \otimes
\frac{\partial H({\bf k},{\bf r};T)}{\partial {\bf
r}_\nu}]        \otimes
\rho({\bf k},{\bf r};\omega,T) \} 
\nonumber \\
&& {\bf J}^{\bf z}({\bf r}, T)=
\int \frac{d\omega}{4 \pi} \frac{d^3{\bf k}}{(2\pi)^3}
Tr\{ {\bf n}({\bf k}) \cdot {\bf \sigma}
[ v_k{\bf \Omega}_{\mu}({\bf k}) 
+\Sigma^K_{\mu\nu}({\bf k}) \otimes
\frac{\partial H ({\bf k},{\bf r};T)}{\partial {\bf
r}_\nu} ] \otimes      
\rho ({\bf k},{\bf r};\omega,T) \}
\label{cstoy}
\end{eqnarray}
Eq.\ref{cstoy} can be used to analyze contributions to the spin and
charge pumping currents from different part of Fermi surfaces. 

Let us define {\em plus} and {\em minus} Fermi seas as shown schematically in Fig.2.
In the {\em plus} Fermi sea, electron spins are along the direction
of unit vector ${\bf n}({\bf k})$ and in the minus Fermi sea spins are along
the opposite direction of ${\bf n}({\bf k})$. 
The corresponding Fermi surfaces are named as the {\em plus} and {minus} fermi surfaces.
For a system with an inversion symmetry each fermi sea has zero overall polarization when the Zeeman field 
is absent. In the rotated basis, these two fermi seas correspond to spin-up and spin-down
ones.

Consider electrons subject to pumping potential gradient along the negative $x$-axis.
One easily finds that spin-up
electrons at the north pole of {\em plus} Fermi
surface
are subject to a drift along the y-direction 
while spin-down electrons  
at the south pole of {\em plus} fermi surface are subject to a drift along the 
minus $y$-direction.
For the same reason electrons at the north and south pole of {\em minus} 
Fermi surface are subject to a drift along {\em minus} and {\em plus} $y$ direction 
respectively (see Fig.4c).

Following these results one also finds that in the absence of spin 
polarization, charge pumping currents carried by spin-up electrons at either the 
north pole of
{\em plus} Fermi surface or the south pole of {\em minus} Fermi surface flow
in the exactly opposite 
direction of
charge pumping currents carried by spin-down electrons in the south
pole of {\em plus} Fermi surface and the north pole of {\em minus} Fermi 
surface.
So while the spin pumping current flows, the net charge pumping vanishes.
Only when electrons are polarized or ${\bf B}$ is nonzero, charge pumping is possible along the
transversal direction.

Given the expressions for $\rho^A$ in Eq.\ref{asy} and the current 
expressions in appendix B, one evaluates
the transverse charge and spin pumping currents,

\begin{eqnarray}
&& {\bf J}_\mu({\bf r})=
\int \frac{d\omega}{2\pi} 
\frac{d^3{\bf k}}{(2\pi)^3}
\frac{1}{D_k} Tr\{
\Sigma^{K}_{\mu\nu}({\bf k}) \otimes
\frac{\partial H ({\bf k}, {\bf r};T)}
{\partial {\bf r}_\nu}
M^{S1}({\bf k},{\bf r};\omega,T) \}
%[I_0 +
%G({\bf X},{\bf X}')
%\frac{\partial H({\bf k}, {\bf X}';T)}{\partial T}
%\frac{\partial}{\partial \omega} ] \otimes
%\frac{\partial}{\partial \epsilon_k} 
%\rho^0({\bf k};\omega)
\\ && 
{\bf J}^z_\mu({\bf r})=\int \frac{d\omega}{4\pi} 
\frac{d^3{\bf k}}{(2\pi)^3} \frac{1}{D_k} Tr\{ {\bf n}({\bf k}) \cdot \sigma
\Sigma^{K}_{\mu\nu}({\bf k})\otimes
\frac{\partial H({\bf k}, {\bf r};T)
}{\partial 
{\bf r}_\nu} 
M^{S1}({\bf k},{\bf r};\omega,T)\}.
\label{TMK}
\end{eqnarray}

Taking into account Eq.\ref{monopole},
one then arrives at expressions for charge and spin pumping currents. 
The longitudinal spin and charge pumping are still given by
Eq.\ref{lcp1}.  The transverse spin and charge pumping currents are more 
involved.
To evaluate the spin and charge current, one notices the following
identities

\begin{eqnarray}
&& Tr(\sigma^z \Sigma_{xy})=2\Sigma^{Kz},
Tr(\sigma^x \Sigma^K_{xy})= Tr(\sigma^y \Sigma^K_{xy})=0
\end{eqnarray}
according to Eq.\ref{monopole}.
Final 
expressions
for spin and charge pumping currents only depend on the $z$-
component of reduced SU(2) fields.
For this reason, one is able to obtain
a rather simple form for transversal adiabatic curvatures 
defined in Eq.\ref{tcp1}

\begin{eqnarray}
&& \Pi^{ts}=\int {d\epsilon_{\bf k}}\int 
\frac{d{\bf \Omega}({\bf k})}{4\pi}
\frac{1}{D_k} {\bf n}_z({\bf k}) \Sigma^{Kz}_{xy} 
\frac{\partial 
\nu(\epsilon_{\bf 
k})}{\partial \epsilon_{\bf k}} n_0(\epsilon_{\bf k});
\nonumber\\
&& \Pi^{tc}=\int {d\epsilon_{\bf k}}\int 
\frac{d{\bf \Omega}({\bf k})}{4\pi} \frac{2}{D_k}
g\mu_B \Gamma \rho_{\bf k} 
[{\Omega}_x^2+{\Omega}_y^2+(\Omega_z + I(\rho_{\bf k}))^2]^{1/2}
\Sigma^{Kz}_{xy} 
\frac{\partial 
\nu(\epsilon_{\bf k})}{\partial \epsilon_{\bf k}} 
\frac{\partial  
n_0(\epsilon_{\bf k})}{\partial \epsilon_{\bf k}} 
\end{eqnarray}

The {\em topological motor} $\Sigma^{Kz}_{\mu\nu}$ in this case is an anisotropic monopole
discussed in Eq.\ref{monopole}.
I only present results in the following weak Zeeman field and strong Zeeman field limits.

Since the topological fields have distinct large momentum and small momentum
asymptotics, the topological pumping has strong dependence on Zeeman 
fields. 
The spin and charge topologically pumped out per period are again given 
by Eq.\ref{tcp1};
the transverse adiabatic curvatures $\Pi^{tc,ts}$ are calculated 
and results are

\begin{eqnarray}
&& \Pi^{ts}= 
\pi^{ts}(\frac{k_F}{\lambda_m})\frac{1}{\pi^{lc}}
\frac{1}{D_0 m} 
\Pi^{lc}; \nonumber\\
&&
\Pi^{tc}=\pi^{tc} (\frac{k_F}{\lambda_m})\frac{1}{\pi^{lc}}
\frac{1}{D_0 m}\frac{g\mu_B B}{\epsilon_F} 
\Pi^{lc}.
\end{eqnarray}
$\Pi^{lc},\pi^{lc}$ are provided in section III; $D_0$ is the diffusion constant and
$m$ is the electron mass. 
I should mention that when the Zeeman field vanishes,
$\Pi^{tc}$ goes to zero but $\Pi^{ts}$ remains finite; 
as pointed out before, this is a distinct
feature of a topological spin pump where an electron beam splits because of
{\em TSGS}.

{\bf $B \ll \Gamma {k_F}$}
This corresponds to a limit where the core size of monopole 
$\lambda_m$ is much smaller than $k_F$,

\begin{eqnarray}
&&
\pi^{ts}(\frac{k_F}{\lambda_m})
=\frac{1}{12}\ln \frac{k_F}{\lambda_m},  \pi^{tc}(\frac{k_F}{\lambda_m})
=o(\frac{\lambda_m}{k_F}).
\end{eqnarray}

{\bf $B \gg \Gamma {k_F}$}
This corresponds to a limit where the core size of monopole 
$\lambda_m$ is much larger than $k_F$,

\begin{eqnarray}
&&
\pi^{ts}(\frac{k_F}{\lambda_m})
=\frac{1}{2} \pi^{tc}(\frac{k_F}{\lambda_m})
=\frac{1}{4} (\frac{k_F}{\lambda_m})^2.
\end{eqnarray}
In deriving these results for $\pi^{ts,tc}$, I have neglected
the k-dependence in diffusion constants and derivatives of
density of states.

The corresponding charge and spin pumping angles are

\begin{eqnarray}
&& \tan \theta_C=
\pi^{tc}(\frac{k_F}{\lambda_m})\frac{1}{\pi^{lc}}
\frac{g\mu_B B}{\epsilon_F} 
\frac{1}{D_0 m}, \nonumber\\
&& \tan \theta_S=
\pi^{ts}(\frac{k_F}{\lambda_m})\frac{1}{\pi^{lc}}
\frac{1}{D_0 m}.
\label{theta}
\end{eqnarray}
Finally, the transversal spin pumping efficiency is 

\begin{eqnarray}
\epsilon^t=\frac{\tan \theta_S}{\tan \theta_C}=
\frac{
\pi^{ts}(\frac{k_F}{\lambda_m})
}{\pi^{tc}(\frac{k_F}{\lambda_m})}
\frac{\epsilon_F}{g \mu_B B}.
\label{epsilont}
\end{eqnarray}

\subsection{The 2D Rashba Model}

In the previous section, I discuss the topological spin pumping due to the ${\bf k}$-space spin rotation in an 
artificial model.
Now I turn to more realistic models for semiconductors.
And I limit ourselves to 2D cases.
Spin-orbit coupling can be either due to the Dresselhaus term or
the Rashba term
\cite{Dresselhaus55,Rashba60,Rashba61,Bychkov84}.
In the later case, or in 
the 
Rashba model for 2D semiconductors,
the spin-orbit coupling has a particularly simple form
because of either a bulk inversion asymmetry or a structure 
inversion asymmetry\cite{Rashba03}. I start with discussions about this model.

In the 2D Rashba model, the spin dependent Hamiltonian can be presented as

\begin{eqnarray}
{H}_{R}=-g\mu_B B_0({\bf k}) [ I({\bf 
k}) {\bf \sigma}^z+  {\bf e}_z 
\cdot {\bf \Omega}\times 
{\bf \sigma}].
\end{eqnarray}
In bracket $[ ]$, the first term is due to
Zeeman fields and the second one is the Rashba coupling term.
As in Eq.65, I have defined $I({\bf k})=B/B_0(|{\bf k}|)$ and $B_0=\Gamma |{\bf 
k}|$. ${\bf \Omega}({\bf k})$ is a 2D unit vector along the direction of ${\bf k}$, 

\begin{eqnarray}
&& \Omega_x=\cos\phi({\bf k})=\frac{k_x}{|{\bf k}|}, \nonumber\\
&& \Omega_y=\sin\phi({\bf k})=\frac{k_y}{|{\bf k}|}.
\end{eqnarray}
Here $\phi({\bf k})$ is a polar angle of ${\bf k}$ vector in the 2D plane. 

To characterize the spin configurations in $(k_x,k_y)$ plane,
I study the unit vector ${\bf n}({\bf k})$ defined as

\begin{eqnarray}                                                               
&& {\bf n}_x=\frac{-{\Omega}_y}                                                 
{\sqrt{1+ I^2}},                                   
\nonumber  \\                                                                  
&& {\bf n}_y=\frac{{\Omega}_x}{\sqrt{1+
I^2}},     
\nonumber \\                                                                   
&& {\bf n}_z=\frac{I}{\sqrt{1+ 
I^2}}.
\label{n2}                                                                      
\end{eqnarray}  
One also obtains simple results for spherical angles 
$\tilde{\theta}({\bf k})$,$\tilde{\phi}({\bf k})$ of 
${\bf n}({\bf k})$ in this case,

\begin{eqnarray}
&& \tilde{\phi}=\phi({\bf k})+\frac{\pi}{2}, \nonumber \\
&&
\tilde{\theta}=\arctan 
\frac{1}{I(|{\bf k}|)}. 
\end{eqnarray}

One notices that at ${\bf k}=0$, $I({\bf k})$ becomes infinity; as a result 
unit vector ${\bf n}({\bf k})$ points at the direction of ${\bf e}_z$ 
because of 
Zeeman 
fields. At the large ${\bf k}$ limit $\tilde{\theta}$ approaches $\pi/2$; 
${\bf n}$ relaxes and lies in the equator plane of two sphere $S^2$. So 
$\tilde{\theta}$ varies
from $0$ to $\pi/2$ as one moves away from the center of Fermi seas 
while $\tilde{\phi}=\phi$. This behavior of
unit vector ${\bf n}$ implies a meron or half-skyrmion in the 2D momentum space. 
Merons
have been proved to
play important roles in Yang-Mills theory as well as in quantum magnetism 
\cite{Gross78,Affleck86}.
The size of half-skyrmion outside which the spin polarization along $z$-direction becomes 
unsubstantial is

\begin{equation}
\lambda_s =\frac{B}{\Gamma}.
\end{equation}

To confirm the peculiar topology of electron spin states one
examines the homotopy class of mapping from $(k_x,k_y)$ space to
a two sphere $S^2$ defined by ${\bf n}({\bf k})$. Consider ${\bf T}^K$ 
defined in Eq.\ref{pyf} in terms of
${\bf n}({\bf k})$ vector instead of ${\bf \Omega}$. The winding number can be easily calculated 

\begin{equation}
W=\frac{1}{2\pi}\int dS {\bf e}_z \cdot {\bf T}^K 
=\frac{1}{2}\int^{\pi/2}_0 d\tilde{\theta} 
\sin\tilde{\theta}=\frac{1}{2}
\end{equation}
which precisely shows a meron in $k_x-k_y$ plane.

In general, $\Gamma$ is a quantity which can be controlled by an electric 
field\cite{Bychkov84}.
$\lambda_s$ is a function of both Zeeman fields and applied external gate voltages which
offers great opportunities to manipulate the meron structure and 
control topological spin pumps
discussed below. In this model,
the adiabaticity condition in Eq.\ref{ad2} requires that
for each ${\bf k}$,

\begin{eqnarray}
&& |\frac{\partial V_{ext}({\bf r})}{\partial \bf r}|
min\{\frac{1}{\lambda_s},\frac{1}{|{\bf k}|} \},
|\frac{\partial V_{im}({\bf r})}{\partial \bf r}|
min\{\frac{1}{\lambda_s},\frac{1}{|{\bf k}|} \}
\ll g\mu_B max\{B_0(|{\bf k}|), B\}. \nonumber \\
\label{ad4}
\end{eqnarray}
Here $V_{im}$ is the impurity optential.
The sufficient condition for Eq.\ref{ad4} to hold is that

\begin{equation}
|\frac{\partial V_{ext}}{\partial \bf r}|,
|\frac{\partial V_{im}}{\partial \bf r}|
\ll \frac{(g\mu_B B)^2}{\Gamma}.
\label{ad5}
\end{equation}
Furthermore, the size of systems has to be larger than
$min\{\lambda_s^{-1}, |{\bf k}|^{-1}\}$ to ensure the semiclassical approximation\cite{limit}.

At last I would like to emphasize one more time that
the impurity potential has to be weak compared with either the Zeeman splitting or the splitting due to spin-orbit coupling so that the 
adiabaticity condition can be satisfied.
Therefore, the transitions 
between the split spin bands are also negligible in the adiabatic limit.
As mentioned in a few occasions in this article, for this reason the Born approximation is always valid in the adiabatic approximation\cite{Zhou04}.

Similar to the procedure introduced in the previous section,
it is possible to
introduce spin rotation to diagonalize this Hamiltonian.
In the rotated basis, spin-up and spin-down states are split by
a combined field of external Zeeman splitting and internal spin-orbit 
coupling with the following strength, $g\mu_B B_0 \sqrt{1 +I^2(|{\bf 
k}|)}$.
The resultant $U(1)$ gauge fields
in the two-dimension ${\bf k}$-space are vortex-like. I present the result
in polar coordinates ${\bf k}=(\rho_{\bf k},\phi)$;

\begin{eqnarray}
&& {\bf A}^{Kz}_{\phi}({\bf k})=
\frac{1}{2\rho_{\bf k}}\frac{I}{\sqrt{1+I^2}}; \nonumber \\ 
&&\Sigma^{Kz}_{\rho\phi}
=-\frac{1}{2\rho_{\bf 
k}}\frac{1}{(1+I^2)^{3/2}}\frac{\partial 
I(\rho_{\bf k})}{\partial \rho_{\bf k}}.
\label{tf2}
\end{eqnarray}

The asymptotics for the vector field ${\bf T}^K$ are
\begin{equation}
{\bf T}^K({\bf k})=\left\{\begin{array}{cc}
\frac{\lambda_s}{2 \rho^3_{\bf k}}{\bf e}_z,
&  \mbox{when $\rho_{\bf k} \gg \lambda_s$;} \\
\frac{1}{2 \lambda_s^2}(1-\frac{3\rho^2_{\bf k}}{2\lambda_s^2}){\bf e}_z, 
&  \mbox{when $\rho_{\bf k} \ll \lambda_s$.} \\
\end{array} \right .
\end{equation}

In the absence of Zeeman fields, I find $\tilde{\theta}=\theta$ and $\Sigma^{Kz}$ is zero everywhere in
the momentum space except at ${\bf k}=0$. That is

\begin{equation}
\Sigma^{Kz}_{xy}=\pi \delta({\bf k}).
\end{equation}
Note that the topological fields though zero everywhere are singular at the 
origin of the ${\bf k}$-space.
In general, topological fields are 
negligible when $\rho_{\bf k}$ is much larger than 
the skyrmion size $\lambda_s$. However,
I find this is sufficient to produce a 
spin pumping current even if the skyrmion size $\lambda_s$ 
is much smaller than the Fermi radius $k_F$.
The topological motor in this model is a positive meron (see Fig.3).

Let us again define {\em plus} and {\em minus} Fermi seas.
In the {\em plus} Fermi sea, all electron spins are along ${\bf n}({\bf k})$ while in the 
{\em minus} Fermi sea all electron spins are along $-{\bf n}({\bf k})$ direction.
In the rotated basis, the {\em plus} and {\em minus} 
fermi seas become spin-up and spin-down fermi seas respectively.
In {\em plus} and {\em minus} Fermi seas, 
electrons subject to external pumping fields along
$x$-axis drift along plus and minus $y$-direction
respectively, as shown in Fig.4 (d).

The spin and charge topologically pumped out per period are again given 
by Eq.\ref{tcp1};
the transverse adiabatic curvatures $\Pi^{tc,ts}$ are 
given in the following equations

\begin{eqnarray}
&& \Pi^{ts}_R=\int {d\epsilon_{\bf k}}\int 
\frac{d{\phi}({\bf k})}{2\pi}
\frac{1}{D_k}
{\bf n}_z({\bf k}) \Sigma^{Kz}_{yx} 
\frac{\partial 
\nu(\epsilon_{\bf 
k})}{\partial \epsilon_{\bf k}} n_0(\epsilon_{\bf k});
\nonumber\\
&& \Pi^{tc}_R=\int {d\epsilon_{\bf k}}\int 
\frac{d{\phi}({\bf k})}{2\pi} 
2 g \mu_B \Gamma \rho_{\bf k} 
\sqrt{1+I^2({\rho_{\bf k}})}
\frac{1}{D_k}
\Sigma^{Kz}_{yx} \frac{\partial 
\nu(\epsilon_{\bf k})}{\partial \epsilon_{\bf k}} 
\frac{\partial n_0(\epsilon_{\bf k})}{\partial \epsilon_{\bf k}}. 
\end{eqnarray}
I have used subscript {\em R} to refer to the adiabatic curvatures in the Rashba model. 
Taking into account the profile of topological fields in Eq.\ref{tf2},
I obtain the following results for the transverse adiabatic curvatures

\begin{eqnarray}
&& \Pi^{ts}_R= 
\pi^{ts}(\frac{k_F}{\lambda_m})\frac{1}{\pi^{lc}}
\frac{1}{D_0 m} 
\Pi^{lc};\nonumber\\
&&
\Pi^{tc}_R=\pi^{tc} (\frac{k_F}{\lambda_m})\frac{1}{\pi^{lc}}
\frac{1}{D_0 m}\frac{g\mu_B B}{\epsilon_F} 
\Pi^{lc}.
\end{eqnarray}
(see section III for discussions on $\Pi^{lc},\pi^{lc}$).
$\pi^{ts}$ and $\pi^{tc}$ are calculated using
rescaled parameters introduced in Eq.\ref{scale},

\begin{eqnarray}
&& \pi^{ts}(b)
=\frac{1}{2}\int da \frac{\partial e(a)}{\partial a} \frac{f(a)}{h(a)}
\frac{1}{a}\frac{I(ba)}{(1+I^2)^2}\frac{\partial I(ba)}{\partial a} n_0, 
\nonumber\\
&& \pi^{tc}(b)=
\int da  \frac{f(a)}{h(a)}
\frac{1}{a}\frac{I(ba)}{1+I^2}\frac{\partial I(ba)}{\partial a} 
\frac{\partial n_0}{\partial a}.
\end{eqnarray}

I again present results in the limit of strong and weak Zeeman fields.

{\bf $B \ll \Gamma {k_F}$}
This corresponds to a limit where the core size of meron 
$\lambda_s$ is much smaller than $k_F$.

\begin{eqnarray}
&& \pi^{ts}(\frac{k_F}{\lambda_s})
= \frac{1}{2}\pi^{tc}(\frac{k_F}{\lambda_s})=\frac{1}{4}.
\end{eqnarray}

{\bf $B \gg \Gamma {k_F}$}
This corresponds to a limit where the core size of meron 
$\lambda_s$ is much larger than $k_F$.

\begin{eqnarray}
&&\pi^{ts}(\frac{k_F}{\lambda_s})
=\frac{1}{2}\pi^{tc}(\frac{k_F}{\lambda_s})
=\frac{1}{4} (\frac{k_F}{\lambda_s})^2.
\end{eqnarray}

The corresponding charge and spin pumping angles,
the transversal spin pumping efficiency are still given by Eq.\ref{theta},
\ref{epsilont}; 
for the Rashba model, 
$\pi^{ts}$ and $\pi^{tc}$ 
calculated above should be used to determine the angles and efficiency.
Without losing generality, I again have neglected the k-dependence 
in $D_k$ and
$\partial \nu(\epsilon_{\bf k})/\partial \epsilon_{\bf k}$ 
in deriving results for $\pi^{ts,tc}$ in this section;
in the 2D model, I further choose to work in a limit where 
$\partial \nu(\epsilon_{\bf k})/\partial \epsilon_{\bf k}$ is nonvanishing because 
of band structures so that the longitudinal charge pumping is nonzero.

\subsection{The 2D Dresselhaus Model}

In the Dresselhaus model, the spin-orbit coupling and Zeeman coupling are given
as

\begin{equation}
H_{D}=- g\mu_B B_0 [{\bf \sigma}\tilde{\cdot} \Omega({\bf k})
+\sigma_z I(|{\bf k}|)];
\end{equation}
here ${\bf A} \tilde{\cdot}{\bf B}$ is defined as $A_x B_x-A_y B_y$.
Again one introduces a unit vector ${\bf n}({\bf k})$ 
to specify spin configurations in fermi seas.
The unit vector ${\bf n}({\bf k})$ characterized by $\tilde{\theta}
({\bf k})$ and $\tilde{\phi}({\bf k})$ is given in terms of $I(|{\bf k}|)$ 
and 
$\phi$ in the following equations,

\begin{eqnarray}
&& \tilde{\phi}=-\phi({\bf k}),\nonumber \\
&& \tilde{\theta}=\arctan
\frac{1}{I(|{\bf k}|)}.  
\end{eqnarray}

Calculations for spin pumping current are identical to
those in the previous section. The winding number of the configuration defined by ${\bf n}({\bf k})$ in this case
is

\begin{equation}
W=\frac{1}{2\pi}\int dS {\bf e}_z\cdot {\bf T}^K
=-\frac{1}{2}\int^{\pi/2}_0d\tilde{\theta}\sin\tilde{\theta}=-\frac{1}{2}
\end{equation}
representing a meron with negative one half skyrmion charge.
This is topologically distinct from the spin configuration in the Rashba 
model. Naturally,
all topological fields in this model are pointing in the negative $z$-axis.

In the Rashba model I find that the {\em topological motor} is a positive
meron in fermi seas while in the Dresselhaus model the {\em motor} is a negative
meron. I anticipate that spins
should be pumped out in an opposite transverse direction in these two models.
This is as well true for  
charge pumping currents in two limits when a 
Zeeman field is present.
So both topological spin and charge pumping currents flow
in an opposite transverse direction compared with currents 
in the Rashba model.

More specifically, I find that
transverse adiabatic curvatures in the Dresselhaus Model (I use subscript
{\em D} to refer to this model) are related to those in the Rashba model (I use subscript {\em R} for this model)
via the following identity
\begin{equation}
\Pi^{ts,tc}_D= - \Pi^{ts,tc}_R.
\end{equation}
This is an exact result as far as the adiabaticity conditions in Eq.\ref{ad1},\ref{ad2}
are satisfied and is independent of the spin-orbit coupling strength in the 2D Rashba and
Dresselhaus models.

In all models, I have found that topological spin and charge pumping
are suppressed by strong Zeeman fields because of spin polarization.
This is a general feature of topological pumps;
the topological fields are absent when electrons are completely Zeeman 
polarized.
Furthermore, the topological charge pumping has a maximum
when the Zeeman field is comparable to spin-orbit fields, i.e. $B\sim 
\Gamma k_F$.
In the appendix C, I have found similar effects on the spin Hall and Hall
conductivity.
In the case where both the Rashba term and Dresselhaus term
are present and controllable, it is interesting to understand 
how a topological 
pump reverses the direction of its spin flow.

\section{Collective responses of various Fermi seas to pumping potentials: transversal displacement versus 
self-twist}

I also would like to emphasize that different schemes of spin pumps discussed 
in this article correspond to 
different responses of Fermi seas to adiabatic perturbations.
To study displacement of Fermi seas or more general deformations of
various Fermi seas, one examines the adiabatic displacement when 
pumping fields are applied along
the $x$-axis. 
It is useful to introduce the displacement vector-tensor ${\bf d}({\bf 
k})$
to characterize collective motion of fermi seas,
\begin{eqnarray}
&& \delta \rho ({\bf k},{\bf r};T)=m {\bf v} \cdot {\bf d} ({\bf k})
\frac{\partial \rho^0 ({\bf k})}{\partial \epsilon_k},
\nonumber \\
&& \delta \rho ({\bf k},{\bf r};T)=\int \frac{d\omega}{2\pi} [\rho ({\bf 
k},{\bf 
r};\omega,T)-\rho^0 ({\bf k};\omega)], \rho^0 ({\bf k})=\int 
\frac{d\omega}{2\pi} \rho^0 ({\bf k};\omega).
\end{eqnarray}

In the adaibatic approximation employed in this article,
only the diagional part of 
displacement tensor is nonzero. One studies these elements 
to analyze the responses of Fermi seas to pumping fields.
For weakly polarized electrons in orbital magnetic fields, spin-up and spin-down Fermi 
surfaces are displaced in the same longitudinal and transverse directions; 
electrons in each Fermi sea drift with almost the same velocity when the Zeeman splitting is much smaller
than the Fermi energy.
Following the calculations in section II,

\begin{equation}
{\bf d}_{\mu}=-\frac{\tau_0}{m} \frac{\partial V_{ext}({\bf r},T)}{\partial 
{\bf 
r}_\nu}\delta_{\nu x}
[\delta_{\mu\nu}+\tau_0 \Omega_c (1+\sigma^z\frac{g\mu_B B}{\epsilon_F}) 
\epsilon_{z \nu\mu}]
\end{equation} 
The responses of Fermi surfaces are summarized in FIG.4.

In an {\em XSSR} based pump, the
{\em plus} and {\em minus} fermi surfaces are displaced along the same 
longitudinal direction but along opposite transverse $v_y$-directions.

\begin{equation}
{\bf d}_{\mu}=-\frac{\tau_0}{m} \frac{\partial V_{ext}({\bf r},T)}
{\partial {\bf r}_\nu}
\delta_{\nu x}[\delta_{\mu \nu}  +\sigma^z 
\frac{\tau_0}{m} {\Sigma^{Xz}_{\mu\nu}}].
\end{equation}

For {\em KSSR} based pumping, Fermi seas respond 
in very distinct ways.
In the generalized Luttinger model, the Rashba model and the Dresselhaus model,
Fermi seas only experience displacement in the 
longitudinal direction;
\begin{equation}
{\bf d}_{\mu}=-\frac{\tau_0}{m} \frac{\partial V_{ext}({\bf r},T)}{\partial 
{\bf r}_\nu}
\delta_{\mu \nu}\delta_{\nu x}
\end{equation}
And Fermi seas are not displaced along the transverse $v_y$-direction.
In this regard, there is a fundamental difference between
spin pumping of polarized electrons, an {\em XSSR} spin pump and
a {\em KSSR} spin pump. In the later case, the one-particle density 
matrix, 
surprisingly,
doesn't develop an asymmetric component along the transverse direction.
The spin current therefore has characteristic of persistent currents which doesn't
involve distortion of Fermi seas. This observation appears to 
be consistent with a proposed analogy between a spin injection current in the Luttinger Model and a supercurrent  
in Ref.\cite{Murakami03a}. 

However, in this case the group velocity of electrons
acquires a transverse term in the rotated basis along the transverse $v_y$-direction;
the dispersion of group velocity tensor is 
\begin{eqnarray}
&& {\bf v}_{\mu}=
v{\bf \Omega}_{k\mu} 
+\sigma^z \Sigma^{Kz}_{\mu\nu}({\bf k})
\frac{\partial V_{ext}({\bf r})}{\partial {\bf       
r}_\nu}.
\label{gv}
\end{eqnarray}
Consider the toy model in section VI A,
in terms of group velocities
one finds that the upper half of the {\em plus} Fermi sea is displaced 
along the positive $v_y$-direction and
the lower half twists along the minus $v_y$-direction.
For the {\em minus} Fermi sea the upper half twists along minus $v_y$ direction
and the lower half twists along the plus $v_y$-direction.
Therefore {\em plus}, {\em minus} Fermi surfaces experience two distinct self-twists.
All these occur while there is no displacement of fermi seas along the $v_y$-direction.

This is as well true for the Rashba model and Dresselhaus model. The 
displacement vector tensor doesn't have a 
transverse component; only the group velocities of electrons 
develop a $v_y$-component. Following Eq.\ref{gv}, the dispersion of group velocity
is given by the calculated meron fields ${\bf T}^{K}$ in section VI C.

So I find that in an {\em XSSR} spin pump, Fermi seas 
experience
periodical displacement along the $v_y$-direction and at any moment, spin-{\em plus}
and spin-{\em minus}
fermi seas are displaced towards two opposite points in the $v_y$ axis.
In the {\em KSSR} pumping scheme on the other hand, Fermi 
seas 
do not have periodical transverse displacement; however group velocities
are renormalized by external fields.  
These responses of Fermi seas lead to topological spin pumping phenomena.

\section{conclusion}

To summarize, I have developed a kinetic approach to analyze topological 
spin pumps in some details.
In all topological spin pumps discussed here, spins as well as charges are 
pumped out in a transversal direction by various topological spin 
configurations which
I have called {\em topological motors}.
I have introduced the spin pumping efficiency and spin pumping angle
to characterize topological spin pumps which could operate 
in the absence of spin polarization through a mechanism of {\em TSGS}.

Following Eqs.63,91
for a given longitudinal charge pumping current, the transversal 
spin pumping angle for
{\em XSSR} based pumps $\theta_S$ 
is bigger in clean systems than in dirty systems; this is similar to the usual Hall angle in dirty metals. 
Transversal spin pumping in this case is attributed to distinct
displacement of spin-{\em plus} and spin-{\em down} fermi seas.  
On the other hand, following Eq.91
for {\em KSSR} based pumping, the transversal spin pumping angle $\theta_S$ is
bigger in dirty structures.
Transversal topological spin pumping in this case occurs when Fermi seas 
are not displaced along the transverse direction.

A unique feature of topological pumps is its Zeeman field dependence.
We have illustrated that when Zeeman fields are much stronger than 
spin-orbit coupling fields, the topological pumping mechanism 
is strongly suppressed.
In the 2D Rashba model, a topological spin pump is driven by a meron
living in Fermi seas while in the Dresselhaus model, spins are pumped out
in a transversal direction by a meron with a negative half skyrmion charge.
The suppression 
occurs as the size of meron increases.
This general feature also appears in the spin Hall and Hall currents in these models. 
In a subsequent paper, I am going to discuss a design
which can be used to establish  
a topological spin pump in laboratories.

Finally I would like to point out
that generalization to various mesoscopic quantum systems is possible.
A mesoscopic mechanism of charge pumping was proposed and studied in
a few recent works \cite{Spivak95,Brouwer98,Zhou99}.
The possibility of pumping charges in an open mesoscopic structure was first noticed 
in Ref.\cite{Spivak95}.
In an extreme quantum 
limit, external 
potentials can be
applied to manipulate coherent wave packets rather than to vary chemical potentials;
the resultant pumping current therefore is greatly 
enhanced at low temperature limit and  decays rapidly at a few hundred mk 
to one kelvin range.
Such a quantum electron pump has been achieved in a remarkable experiment\cite{Marcus99}.

A few interesting aspects of coherent charge pumping were addressed in 
later works.
The symmetry of charge pumping in the quantum limit was studied in Ref. \cite{Shutenko00}. 
The issue of counting statistics was raised and addressed in a fascinating work\cite{Levitov01};
the counting statistics might shed light on the issue of dissipation in an adiabatic
pump and remains to be studied in experiments.
Pumping of coherent Andreev states was analyzed in Ref.\cite{Zhou01}.
At last I should also mention that
various other proposals for spin pumps have been made by different 
groups; 
see for example\cite{Sharma01,Mucciolo02,Zheng02,Wu02,Watson03,Sharma03}.
I would like to refer 
readers to those works for detailed discussions in various limits. 
In chaotic quantum dots where current experiments are carried out, an adiabatic spin pump of coherent 
polarized electrons was proposed in Ref.\cite{Mucciolo02};
the effect of spin-orbit scattering has been further studied 
in a recent article \cite{Sharma03}.

{\bf Acknowledgement}
I want to acknowledge very helpful discussions with I.Affleck, 
M. Berciu, M.Franz, Q. Niu,
G.Sawatsky, B.Spivak and P.E.C.Stamp;
I also would like to thank S. C. Zhang for discussions on 
his recent works and sending me a copy of an early work.
This work was partially supported by the FOM, the Netherlands 
under contract 00CCSPP10, 02SIC25 and by the NWO-MK "projectruimte" 00PR1929; 
it is also supported by a research grant from UBC, Canada.

\newpage
\appendix

\section{Current expressions in the presence of {\em XSSR}}

Under the spin rotation defined in section V,
I would like to underscore the following transformation
in the equation of motion for the one-particle density matrix

\begin{eqnarray}
&& \frac{\partial H ({\bf k},{\bf r};T)}{\partial
{\bf r}_\mu} \rightarrow  \frac{\partial H ({\bf k},{\bf r};T)}{\partial
{\bf r}_\mu}+
\Sigma^X_{\mu\nu}({\bf r})
\otimes
\frac{\partial H({\bf k},{\bf 
r};T)}{\partial {\bf k}_\nu}. 
\end{eqnarray}

Because of gauge fields in the rotated
basis, the 
asymmetric component of the density matrix acquires an additional term 
representing
the transverse motion of electrons in the presence of topological fields.
Calculations show 

\begin{eqnarray}
&&\rho^{A}({\bf k},{\bf r};\omega, T)=
\tau_0 v {\bf \Omega}_{k\mu}
[-\frac{\partial}{\partial {\bf r}_{\mu}}+\frac{\partial 
H({\bf k},{\bf r};T)}{\partial 
{\bf r}_\mu}
\frac{\partial}{\partial \epsilon_k}+ 
D_k \Sigma^X_{\mu\nu}({\bf r}) \otimes \frac{\partial H
({\bf k},{\bf r};T)}{\partial {\bf r}_\nu}
\frac{\partial^2}{\partial^2 \epsilon_k}]\otimes\rho^S
({\bf k},{\bf r};\omega,T).
\end{eqnarray}
The solution of $\rho^S$ is still given by Eq.\ref{asy}.

Taking into account the asymmetric component derived above, I
obtain various charge and spin currents in both longitudinal and 
transverse directions. 
Supercripts $l,t$ 
stand for longitudinal and transverse directions,
$c$ and $s$ stand for charge or spin currents; $1,2$ are for transport and
pumping currents respectively.

Various charge currents are

\begin{eqnarray}
&& {\bf J}^{lc1}_\mu({\bf r})=
\int \frac{d\omega}{2\pi} 
\frac{d^d{\bf 
k}}{(2\pi)^d} D_k Tr \{ 
\frac{\partial H({\bf k},{\bf r};T)}{\partial 
{\bf r}_\mu}\otimes 
\frac{\partial }{\partial \epsilon_{\bf k}}
\rho^{0}({\bf           
k};\omega) \}, 
\nonumber\\
&& {\bf J}^{lc2}_\mu({\bf r})=
\int \frac{d\omega}{2\pi} 
\frac{d^d{\bf 
k}}{(2\pi)^d} D_k Tr \{
\frac{\partial H({\bf k},{\bf r};T)}{\partial 
{\bf r}_\mu}\otimes 
\frac{\partial }{\partial \epsilon_{\bf k}}
[\frac{1}{D_k}M^{S1}({\bf           
k},{\bf r};\omega,T)] \};
\nonumber\\
&&  {\bf J}^{tc1}_\mu({\bf r})=
\frac{\tau_0 }{m}
\int \frac{d\omega}{2\pi} 
\frac{d^d{\bf 
k}}{(2\pi)^d}
D_k Tr\{ [ \Sigma^{X}_{\mu\nu}({\bf r})
\otimes \frac{\partial H({\bf k},{\bf r};T)}{\partial
{\bf r}_\nu}] 
\otimes \frac{\partial }{\partial 
\epsilon_{\bf k}}
\rho^0({\bf k};\omega) \},
\nonumber\\
&&  {\bf J}^{tc2}_\mu({\bf r})=
\frac{ \tau_0 }{m}
\int \frac{d\omega}{2\pi} 
\frac{d^d{\bf 
k}}{(2\pi)^d} D_k Tr\{
\Sigma^{X}_{\mu\nu}({\bf r})
\otimes \frac{\partial H({\bf k},{\bf r};T)}{\partial
{\bf r}_\nu} 
\otimes \frac{\partial }{\partial 
\epsilon_{\bf k}}[\frac{1}{D_k}
M^{S1}({\bf           
k},{\bf r};\omega,T)] \}.
\end{eqnarray}

Spin currents are

\begin{eqnarray}
&& {\bf J}^{z,ls1}_\mu({\bf r})=
\int \frac{d\omega}{4\pi} 
\frac{d^d{\bf 
k}}{(2\pi)^d}
Tr \{ {\bf n}\cdot {\bf \sigma}
D_k \frac{\partial H({\bf k},{\bf r};T)}{\partial 
{\bf r}_\mu}\otimes 
\frac{\partial }{\partial \epsilon_{\bf k}}
\rho^{0}({\bf           
k};\omega) \}, 
\nonumber\\
&& {\bf J}^{z,ls2}_\mu({\bf r})=
\int \frac{d\omega}{4\pi} 
\frac{d^d{\bf 
k}}{(2\pi)^d}
Tr \{ {\bf n} \cdot {\bf \sigma}
D_k
\frac{\partial H({\bf k},{\bf r};T)}{\partial 
{\bf r}_\mu}\otimes 
\frac{\partial }{\partial \epsilon_{\bf k}}[\frac{1}{D_k}
M^{S1}({\bf           
k},{\bf r};\omega,T)] \};
\nonumber\\
&&  {\bf J}^{z,ts1}_\mu({\bf r})=
\frac{ \tau_0 }{m}
\int \frac{d\omega}{4\pi} 
\frac{d^d{\bf 
k}}{(2\pi)^d} D_k  Tr\{
{\bf n}\cdot {\bf \sigma}
[ \Sigma^{X}_{\mu\nu}({\bf r})
\otimes \frac{\partial H({\bf k},{\bf r};T)}{\partial
{\bf r}_\nu} ]
\otimes \frac{\partial }{\partial 
\epsilon_{\bf k}}
\rho^0({\bf k};\omega) \}
\nonumber\\
&&  {\bf J}^{z,ts2}_\mu({\bf r})=
\frac{\tau_0 }{m}
\int \frac{d\omega}{4\pi} 
\frac{d^d{\bf 
k}}{(2\pi)^d} D_k Tr \{
{\bf n} \cdot {\bf \sigma}
\Sigma^{X}_{\mu\nu}({\bf r})
\otimes  \frac{\partial H({\bf k},{\bf r};T)}{\partial
{\bf r}_\nu} 
\otimes \frac{\partial }{\partial 
\epsilon_{\bf k}} [\frac{1}{D_k}
M^{S1}({\bf           
k},{\bf r};\omega,T)] \};
\end{eqnarray}
$d=2,3$ is the dimension of fermi seas.

Currents ${\bf J}^{tc1}$ and ${\bf J}^{z,ts1}$ in these expressions 
yield contributions to
the anomalous
Hall effect and spin Hall current respectively.
Superscript $S1$ has been introduced to specify the symmetric
part of density matrix which is linear in the external frequency.

\section{Current expressions in the presence of {\em KSSR}}

In the presence of $KSSR$,
the longitudinal spin or charge currents are the same as given in the previous 
subsection. The transverse spin and charge transport (1) or pumping (2) 
currents are

\begin{eqnarray}
&& {\bf J}^{tc1}_\mu({\bf r})=
\int \frac{d\omega}{2\pi} 
\frac{d^d{\bf k}}{(2\pi)^d} 
Tr \{ \Sigma^{K}_{\mu\nu}({\bf k}) \otimes
\frac{\partial 
H ({\bf k}, {\bf r};T)}
{\partial 
{\bf r}_\nu}
\rho^0({\bf k} ;\omega)\} \nonumber\\
&& {\bf J}^{tc2}_\mu({\bf r})=
\int \frac{d\omega}{2\pi} 
\frac{d^d{\bf k}}{(2\pi)^d} 
Tr\{ \Sigma^{K}_{\mu\nu}({\bf k}) \otimes
\frac{\partial 
H ({\bf k}, {\bf r};T)}
{\partial 
{\bf r}_\nu}
 \frac{1}{D_k}
M^{S1}({\bf k},{\bf r};\omega,T) \}
\nonumber\\
&& {\bf J}^{z,ts1}_\mu({\bf r})=
\int \frac{d\omega}{4\pi} 
\frac{d^d{\bf k}}{(2\pi)^d} Tr\{
{\bf n}({\bf k}) \cdot \sigma
\Sigma^{K}_{\mu\nu}({\bf k}) \otimes
\frac{\partial 
H ({\bf k}, {\bf r};T)}
{\partial 
{\bf r}_\nu}
\rho^0({\bf k} ;\omega)\}\nonumber\\
&& {\bf J}^{z,ts2}_\mu({\bf r})=
\int \frac{d\omega}{4\pi} 
\frac{d^d{\bf k}}{(2\pi)^d} 
Tr \{ {\bf n}({\bf k}) \cdot \sigma
\Sigma^{K}_{\mu\nu}({\bf k}) \otimes
\frac{\partial 
H ({\bf k}, {\bf k};T)}
{\partial 
{\bf r}_\nu}
 \frac{1}{D_k}
M^{S1}({\bf k},{\bf k};\omega,T) \}.
\end{eqnarray}

\section{Zeeman field dependence of the Hall and spin Hall currents
in the Rashba Model}

In this section I am going to apply results obtained in  Appendix B to
briefly discuss the Hall current and spin Hall current.
The spin hall current in the Rashba Model has been studied in various papers
and I do not intend to reproduce all results in details.
Here I will address the effect of Zeeman fields using the 
approach developed above.

Define the Hall conductivity and spin Hall conductivity
in a usual way,

\begin{equation}
{\bf J}_{y}=\sigma_{yx} {\bf E}_x, {\bf J}^z_{y}=\sigma^z_{yx} 
{\bf E}_x.
\end{equation}
After taking into account the symmetry of $\Sigma^{K}_{xy}$,
I find the following expressions for spin Hall and Hall conductivity
in terms of $\Sigma^{Kz}_{xy}$,

\begin{eqnarray}
&& \sigma_{yx}^z=\frac{1}{2} \int\frac{d^2{\bf k}}{(2\pi)^2}
2{\bf n}_z({\bf k}) \Sigma^{Kz}_{xy}({\bf k})n_0(\epsilon_{\bf k}),
\nonumber\\
&& \sigma_{yx}= \int\frac{d^2{\bf k}}{(2\pi)^2}
2 g\mu_B\Gamma \rho_{\bf k}
[1+I^2({\bf k})]^{1/2}
\Sigma^{Kz}_{xy}({\bf k}) 
\frac{\partial}{\partial \epsilon_{\bf k}} n_0(\epsilon_{\bf k}).
\end{eqnarray}

The rest of calculations is straightforward and  
I derive the following results for the spin Hall and Hall
conductivity in terms of  
$I({\rho_{\bf k}})$

\begin{eqnarray}
&& \sigma^z_{yx}=- \frac{1}{4\pi} \int d{\rho_{\bf k}} 
\frac{I}{(1+I^2)^2}\frac{\partial 
I}{\partial \rho_{\bf k}} n^0(\epsilon) \nonumber \\
&& \sigma_{yx}=-
\frac{1}{2\pi}\int d{\rho_{\bf k}}\frac{g\mu_B B_0(\rho_{\bf k})}
{1+I^2}\frac{\partial I}{\partial 
\rho_{\bf k}}\frac{\partial n_0(\epsilon_k)}{\partial \epsilon_k}
\end{eqnarray}

Finally,
I obtain the Zeeman field dependence of the spin Hall and Hall conductivity,
\begin{eqnarray}
&&\sigma^z_{yx}=\frac{1}{8\pi}
\frac{k_F^2}{\lambda_s^2+k_F^2},\nonumber\\
&& \sigma_{yx}=\frac{1}{4\pi}
\frac{g\mu_B B}{\epsilon_F}\frac{ k_F^2}{\lambda_s^2+k_F^2}.
\end{eqnarray}
${\lambda_s}$ is defined in section VI B.
As mentioned before, I have set $\hbar=e=1$ in this article.

$\sigma^z_{yx}$ at zero Zeeman field
was studied in a few recent works using 
Kubo formula\cite{Sinova03,Schliemann03}.
I find that $\sigma^{z}_{xy}$ is
a smooth monotonous function of Zeeman fields and decreases as fields increase.
$\sigma_{yx}$ has a maximum at $\lambda_s = k_F$. 
Note all results are valid when the adiabaticity conditions in section VI.B are
satisfied. Since the main contribution to the spin Hall current is actually from states close to $|{\bf k}|=0$, it is
essential to lift the degeneracy at ${\bf k}=0$. So 
in the presence of impurity scattering, the adiabaticity condition implies that a finite Zeeman field need to be present
and the limit of zero field should be taken with great care.

\begin{figure}
\begin{center}
\epsfbox{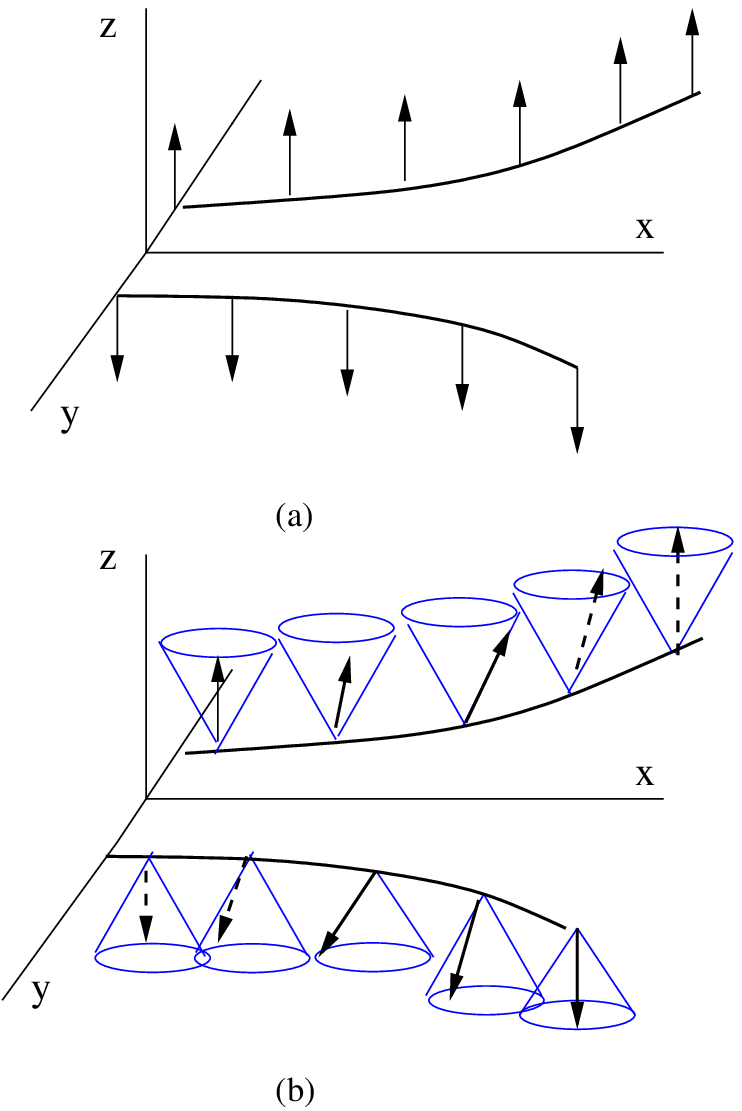}
\leavevmode
\end{center}
\caption{Comparison of the usual Stern-Gerlach beam splitting and
the topological Stern-Gerlach splitting.
When the Zeeman fields are applied along the $z$-direction
as shown in Fig.(a), 
in the conventional Stern-Gerlach splitting 
${\bf S}_z$ is a good quantum 
number and no spin rotation occurs; 
the Zeeman 
field gradient drives spin-up and spin-down particles apart along the y-direction.
However {\em TSGS} is always accompanied by spin rotation as shown
in Fig.(b). Spins are represented by short arrows in this Figure and 
other 
figures. }
\end{figure}

\begin{figure}
\begin{center}
\epsfbox{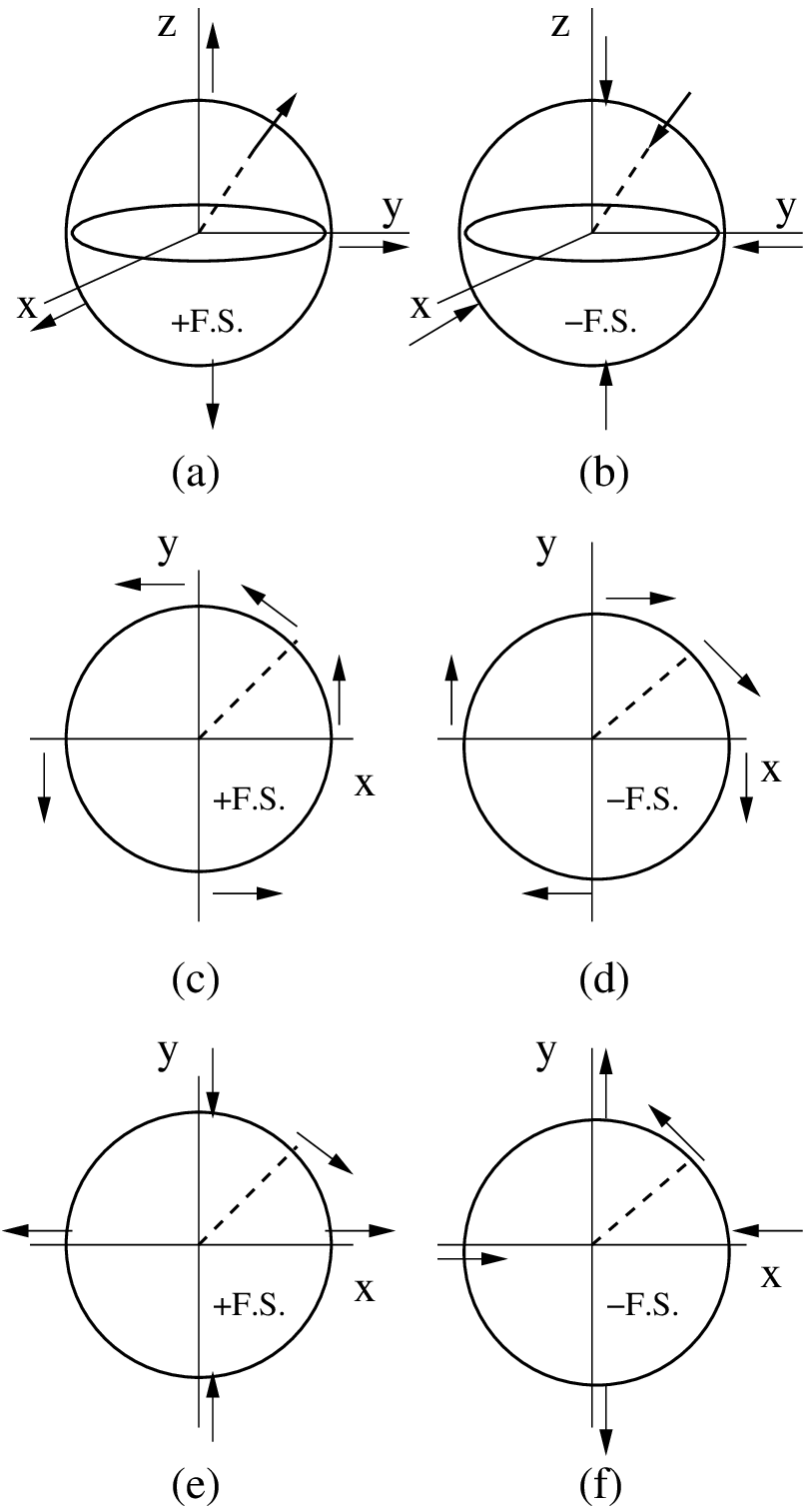}
\leavevmode
\end{center}
\caption{{\em Plus} and {\em minus} Fermi seas for the toy Hamiltonian in 
Eq.\ref{toy}, 
the Rashba Hamiltonian and Dresselhaus Hamiltonian.
Fig.(a),(b) are for the toy Hamiltonian. In the {\em plus} Fermi sea each spin points at the direction of 
its momentum (shown in Fig. (a)); in the {\em minus} Fermi sea shown in Fig.(b), the spin of an electron
points at the opposite direction of its momentum.
In Fig. (c) and (d), we show the corresponding Fermi seas for the 2D Rashba Hamiltonian.
In $k_x-k_y$ plane, electron spins in the {\em plus} fermi seas form a meron with a half skyrmion 
charge when a Zeeman field is applied along the
$z$-axis.
At large ${\bf k}$ limit in the {\em plus} Fermi seas, spins 
point at ${\bf e}\times {\bf \Omega}({\bf k})$ direction while in the 
{\em minus} Fermi seas
spins point at $-{\bf e} \times {\bf \Omega}({\bf k})$; 
they both represent vortices with one unit {\em positive} vorticity.
Here ${\bf \Omega}({\bf k})$ is a unit 
vector along ${\bf k}$. 
At the center of fermi seas, spins are along the $\pm z$
directions (not shown here). 
In Fig.(e) and (f), we show spin rotation in fermi seas of the Dresselhaus model;
electron spins in the {\em plus} fermi seas form a meron with a half negative skyrmion charge.
At large momentum, electron spins in both fermi seas form vortices with one unit {\em negative} 
vorticity. }
\end{figure}

\begin{figure}
\begin{center}
\epsfbox{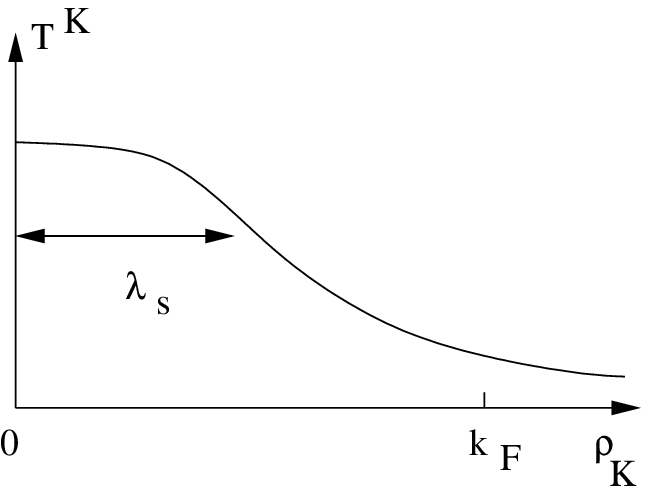}
\leavevmode
\end{center}
\caption{Topological field $T^K$ as a function of $\rho_{\bf k}$ in the 2D Rashba model (schematic).
$\lambda_s$ which defines the size of meron is chosen to be smaller than the fermi momentum $k_F$}
\end{figure}

\begin{figure}
\begin{center}
\epsfbox{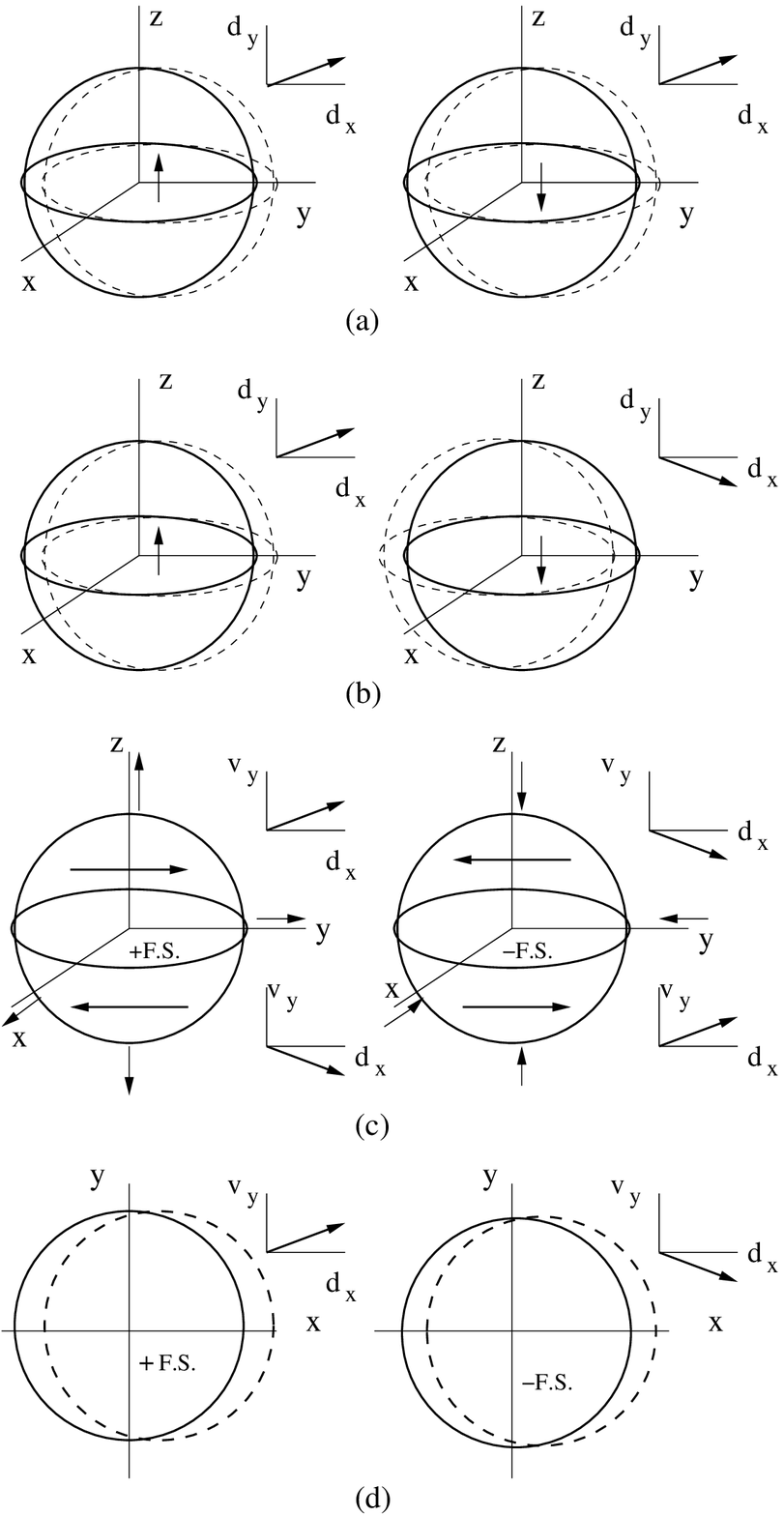}
\leavevmode
\end{center}
\caption{Responses of Fermi seas and individual electrons to external 
pumping fields applied 
along the $x$-direction.   
Fig.(a) is for polarized electrons in the presence of orbital magnetic 
fields directing along the $z$-direction; spin-up and spin-down 
Fermi seas experience {\em identical} displacement ${\bf d}$ along the longitudinal 
direction $x$ and transverse direction $y$ which are shown
in two upper left insets of Fermi seas.
Fig. (b) is for electrons with spin rotation in the $X$-space;
in this case spin-up and spin-down Fermi seas have opposite 
displacement ${\bf d}$
along the transverse direction $y$ as shown in two insets for two Fermi 
seas.
In Fig.(c)
we illustrate differences in electrons' responses in the
{\em plus} and {\em minus} Fermi seas. 
We want to emphasize that
Fermi seas don't have collective displacement along
a transverse direction $y$ or ${\bf d}_y=0$.
In insets we show
the external-field-induced drift or group velocity of individual 
electrons ${\bf v}_y$ in different regions 
of fermi seas.
The transverse drift of an electron in the upper part of a Fermi sea with momentum ${\bf k}$ is in an
opposite direction of 
the drift of the electron with momentum $-{\bf k}$ in the  
lower part, which results in self-twists of Fermi seas.
In upper(lower) insets, we show the drift of electrons in the north (south) poles 
of {\em plus} and {\em minus} fermi seas respectively;
the big arrows across Fermi seas indicate the direction of two distinct twists of {\em plus}
and {\em minus} Fermi seas.
In Fig.(d), we illustrate responses of the {\em plus} and {\em minus} fermi seas in the 2D Rashba model;
in this case, again two fermi seas have zero displacement in the transverse y-directions (${\bf d}_y=0$).
However, the spin-plus and spin-minus electrons in two fermi seas acquire 
a field-induced {\em dispersive} group velocities in the opposite 
$y$-direction as shown in two insets in Fig.(d);
the field-induced transveral group velocity decreases rapidly as the momentum increases.
See section VII for 
detailed discussions.}
\end{figure}

\begin{thebibliography}{99}
\bibitem{limit}This is a stringent sufficient condition. However if the main contribution to currents is 
from electrons with momentum $k_F$ away from the ${\bf k}=0$ point, the condition can be relaxed and is 
always 
satisfied.
\end{thebibliography}
\end{document}